\documentclass[journal]{IEEEtran}
\usepackage{cite}

\ifCLASSINFOpdf
   \usepackage[pdftex]{graphicx}
   \graphicspath{{../pdf/}{../jpeg/}}
   \DeclareGraphicsExtensions{.pdf,.jpeg,.png}
\else
\fi
\usepackage{amsmath}
\usepackage{array}

\ifCLASSOPTIONcompsoc
 \usepackage[caption=false,font=normalsize,labelfont=sf,textfont=sf]{subfig}
\else
 \usepackage[caption=false,font=footnotesize]{subfig}
\fi
\usepackage{bm}
\usepackage{amssymb}
\usepackage{amsthm}
\usepackage[linesnumbered,ruled,vlined]{algorithm2e}
\usepackage{float}
\usepackage{threeparttable}

\newtheorem{theorem}{Theorem}
\newtheorem{corollary}{Corollary}

\newtheorem{proposition}{Proposition}
\newtheorem{definition}{Definition}
\newtheorem{remark}{Remark}
\newtheorem{example}{Example}

\makeatletter
\renewcommand*\env@matrix[1][\arraystretch]{%
	\edef\arraystretch{#1}%
	\hskip -\arraycolsep
	\let\@ifnextchar\new@ifnextchar
	\array{*\c@MaxMatrixCols c}}
\makeatother

\begin{document}
%
\title{On spectral algorithms for community detection in stochastic blockmodel graphs with vertex covariates}
%
%
%


\author{Cong~Mu,
	Angelo~Mele,
	Lingxin~Hao,
	Joshua~Cape,
	Avanti~Athreya,
	and~Carey~E.~Priebe
	\thanks{Date of current version \today.}%
	\thanks{This work was supported in part by the US National Science Foundation under Grant SES-1951005 and in part by the US Defense Advanced Research Projects Agency under the D3M program administered through contract FA8750-17-2-0112.}
	\thanks{C. Mu, A. Athreya, and C.E. Priebe are with the Department of Applied Mathematics and Statistics, Johns Hopkins University, Baltimore, MD 21218. E-mail: \{cmu2, dathrey1, cep\}@jhu.edu.}%
	\thanks{A. Mele is with the Carey Business School, Johns Hopkins University, Baltimore, MD 21202. E-mail: angelo.mele@jhu.edu.}%
	\thanks{L. Hao is with the Department of Sociology, Johns Hopkins University, Baltimore, MD 21218. E-mail: hao@jhu.edu.}%
	\thanks{J. Cape is with the Department of Statistics, University of Pittsburgh, Pittsburgh, PA 15260. E-mail: joshua.cape@pitt.edu.}
	}

\maketitle

\begin{abstract}
	In network inference applications, it is often desirable to detect community structure. Beyond mere adjacency matrices, many real-world networks also involve vertex covariates that carry key information about underlying block structure in graphs. To assess the effects of such covariates on block recovery, we present a comparative analysis of two model-based spectral algorithms for clustering vertices in stochastic blockmodel graphs with vertex covariates. The first algorithm uses only the adjacency matrix, and directly estimates the block assignments. The second algorithm incorporates both the adjacency matrix and the vertex covariates into the estimation of block assignments, and moreover quantifies the explicit impact of the vertex covariates on the resulting estimate of the block assignments. We employ Chernoff information to analytically compare the algorithms' performance and derive the information-theoretic Chernoff ratio for certain models of interest. Analytic results and simulations suggest that the second algorithm is often preferred: one can better estimate the induced block assignments by first estimating the effect of vertex covariates. In addition, real data examples also indicate that the second algorithm has the advantage of revealing underlying block structure while considering observed vertex heterogeneity in real applications. 
\end{abstract}

\begin{IEEEkeywords}
Spectral graph inference, community detection, stochastic blockmodel, vertex covariate, Chernoff ratio.
\end{IEEEkeywords}

%
\IEEEpeerreviewmaketitle

\section{Introduction}

\label{sec:1}

\IEEEPARstart{N}{etwork} data, which encodes interactions or relationships across entities, often involves more than mere links or connections across network vertices. For example, a network dataset may include both an adjacency matrix, which consolidates information about vertices in the network and the edges between them, as well as additional vertex covariates. For example, in  diffusion MRI connectome datasets~\cite{Priebe2019}, vertices represent sub-regions of the brain defined via spatial proximity, and edges represent tensor-based fiber streamlines connecting these sub-regions; such graphs can also have brain hemisphere and tissue labels for each vertex.
Social network datasets~\cite{Leskovec2014,Rozemberczki2019,Rozemberczki2020}, in which vertices can represent users or web pages and edges can represent followers or relationships, may come with ancillary demographic information for each vertex. Since accurate inference on random networks depends on exploiting all available signal, scalable algorithms that can incorporate both network connectivity data and any additional insight from vertex covariates are desirable. For instance, in the well-known $k$-block stochastic blockmodel (SBM)~\cite{Holland1983}, network vertices belong to $k$ distinct groups, or communities, called {\em blocks}, and the probabilities of connection across vertices depend on their block memberships. That is, if $\tau_i$ represents the block associated to vertex $i$, the connection probability between vertex $i$ and $j$ is a function of $\tau_i$ and $\tau_j$. Typically, a vertex's block membership depends on inherent but unobserved (latent) vertex properties. Thus, a classic inference task is to estimate block memberships from a realization of the resulting network. If, however, we observe both adjacency matrices and vertex covariates, and if both can contain information about the latent communities or blocks, we need models and scalable algorithms that can effectively incorporate adjacency structure {\em and} covariate data and account for their potentially disparate effects.

In fact, vertex covariates can influence the very number of communities that are detected in a blockmodel: for example, a two-block SBM might bifurcate further into a four-block SBM because of the impact of a binary covariate (with each block splitting according to the covariate). Standard estimation on a graph, then, may yield a four-block assignment, but understanding the underlying two-block SBM is important in inference applications, as we show. To get to an estimate of the underlying two-block assignment, we need to understand the role of the covariates.

Moreover, a problem of interest in network hypothesis testing is to assess the influence of latent communities on downstream or outcome variables, controlling for vertex covariate effects~\cite{Hao2020,Shalizi2016}. For instance, assume $\mathbf{y}_i$ represents some outcome variable associated to vertex $i$ in a $K$-block SBM (for example, in a demographic data set, $\mathbf{y}_i$ might represent an individual's educational attainment or earnings). Suppose this outcome variable's distribution depends on the vertex's block assignment within the network, so if $\tau_i=k$, then $\mathbf{y}_i$ follows some distribution $F_k$ that can depend on this block. We describe this scenario by writing
$ \mathbf{y}_i | (\tau_i = k) \; \sim \; F_k $. A natural question to ask is whether the distributions of the outcome variables are the same for different blocks, i.e.,~to test whether $ F_k = F $ for $ k \in \{1, \cdots, K \} $. To achieve this goal when we have both adjacency and covariate information, it is crucial to estimate the underlying block structure $ \bm{\widehat{\tau}} $---namely, to obtain an estimate of the block structure {\em after} accounting for, and effectively ``netting out" the vertex covariate effect. Here we write ``induced block assignment'' to refer to the block assignment \emph{after} accounting for the vertex covariates.

As a special case of random graph models, SBMs are popular in the literature for community detection~\cite{Abbe2018,Holland1983,Karrer2011}. Many classical methods consider the adjacency or Laplacian matrices for community detection; see~\cite{Fortunato2016} for an overview. However, these methods are typically not designed to distinguishing the impact of covariates from the mechanism of network generation itself---that is, delineating in the observed data what may be underlying, or fundamental, {\em network} effects from characteristics that are more properly functions of the covariates.  By contrast, covariate-aware inference in SBMs often relies on either variational methods~\cite{Choi2012,Roy2019,Sweet2015} or spectral approaches~\cite{Binkiewicz2017,Huang2018,Mele2019}. For example, \cite{Binkiewicz2017} proposed covariate-assisted spectral clustering (CASC) where the covariates are first parameterized as in linear regression, i.e.,~categorical covariates are represented with dummy variables and continuous covariates can go through standardization, and then combined with the graph for subsequent spectral clustering. The pairwise covariates-adjusted stochastic blockmodel (PCABM), in which pairwise covariate information is incorporated with the classical SBM, was introduced in \cite{Huang2018}. There, model parameters can be solved via maximum likelihood estimation (MLE) or spectral clustering with adjustment (SCWA).

Spectral methods~\cite{Von2007} that promise applicability to large graphs have been widely used in random graph models for a variety of subsequent inference tasks such as community detection~\cite{Lyzinski2014,Lyzinski2016,McSherry2001,Rohe2011}, vertex nomination~\cite{Lyzinski2019}, nonparametric hypothesis testing~\cite{Tang2017}, and multiple graph inference~\cite{Wang2019}. Two particular spectral embedding methods, adjacency spectral embedding (ASE) and Laplacian spectral embedding (LSE), which are spectral decompositions of the graph adjacency and graph Laplacian matrices, respectively, are popular, since they provide consistent~\cite{Sussman2012} and asymptotically normal~\cite{Athreya2016,Tang2018} estimates of underlying graph parameters, such as block memberships. To compare the performance of these two embedding methods, the concept of Chernoff information is employed for SBMs~\cite{Abbe2018,Tang2018} and then extended to consider the underlying graph structure~\cite{Cape2019}. The Chernoff information between two distributions $F_1$ and $F_2$ is related to the exponential rate of decay of the Bayes risk in the simple hypothesis test comparing $F_1$ against $F_2$, as the sample size increases. As such, because of asymptotic normality of the adjacency and Laplacian spectral embedding for stochastic blockmodels, the Chernoff information between two normal distributions (with different mean vectors and covariance matrices) can be adopted to derive the large sample optimal error rate for recovering block assignments in a SBM. 

In this work, we investigate two spectral algorithms for clustering vertices in stochastic blockmodel graphs with vertex covariates. Analytically, we compare the algorithms' performance via Chernoff information and derive the Chernoff ratio  for certain models of interest. The notion of Chernoff information for comparing algorithms will be addressed in detail in Section~\ref{sec:4}. Practically, we compare the algorithms' empirical clustering performance by simulations and real data examples on diffusion MRI connectome and social networks.

The structure of this article is as follows. Section~\ref{sec:2} reviews relevant models for random graphs and the basic idea of spectral methods. Section~\ref{sec:3} introduces our spectral algorithms for clustering vertices in stochastic blockmodel graphs with vertex covariates. Section~\ref{sec:4} analytically compares the algorithms' performance via Chernoff information and derives the Chernoff ratio expression for certain models of interest. Section~\ref{sec:5} provides simulations and real data examples on diffusion MRI connectome and social networks to compare the algorithms' performance. Section~\ref{sec:6} discusses the findings and raises questions for further investigation. Appendices provide technical details for latent position geometry and analytic derivations of the Chernoff ratio as well as the details of simulations. The implementation of our algorithms can be found at https://github.com/CongM/sbm-cov.

\section{Models and Spectral Methods}

\label{sec:2}

To ground our analysis and results, we begin with a particular class of random network models known as latent position models~\cite{Hoff2002,Handcock2007} for edge-independent random graphs. In these models, each network vertex $ i $ is associated with a latent position $ \mathbf{X}_i \in \mathcal{X} $ where $ \mathcal{X} $ is some latent space such as $ \mathbb{R}^d $, and edges between vertices arise independently with probability $ \mathbf{P}_{ij} = \kappa(\mathbf{X}_i, \mathbf{X}_j) $ for some kernel function $ \kappa: \mathcal{X} \times \mathcal{X} \to [0, 1] $.  This is an appealing model to consider not only because of its wide applicability---after all, the kernel can be any reasonable regular function---but because it is easily interpretable. For example, social network connections are often a function of individual participants' (potentially unobserved) interests in a core set of topics or hobbies, and levels of interest can be easily encoded in a low-dimensional space. Moreover, the kernel and this lower-dimensional space can possess intuitive geometry, wherein collinearity or other ``closeness" of latent positions increases the probability of a connection between the associated vertices.  The core model we focus on here, the generalized random dot product graph (GRDPG), has precisely such a property: the kernel function is taken to be the (indefinite) inner product. As the name suggests, this model generalizes the {\em random dot product graph} (RDPG) by relaxing the restriction that the kernel function be the inner product, and this relaxation permits SBM with dissassortative structure, and in fact subsumes all SBMs as special cases.

\begin{definition}[Generalized Random Dot Product Graph~\cite{Rubin-Delanchy2017}]
	\label{def:GRDPG}
	Let $  d = d_+ + d_- $ with $ d_+ \geq 1 $ and $ d_- \geq 0 $. Let $ \mathbf{I}_{d_+ d_-} = \text{diag}\left(1, \cdots, 1, -1, \cdots, -1 \right) $, i.e.,~a $ d \times d $ diagonal matrix with 1 in first $ d_+ $ entries and $-1$ in the next $ d_- $ entries. Let $ \mathbf{A} \in \left\{0, 1 \right\}^{n \times n} $ be an adjacency matrix and $ \mathbf{X} = [\mathbf{X}_1, \cdots, \mathbf{X}_n]^\top \in \mathbb{R}^{n \times d} $ where each $ \mathbf{X}_i \in \mathbb{R}^{d} $ denotes the latent position for vertex $ i $ satisfying $ \mathbf{X}_{i}^\top \mathbf{I}_{d_+ d_-} \mathbf{X}_j \in [0, 1] $ for all $ i, j \in \{ 1, \cdots, n \} $. Then we say $ (\mathbf{A}, \mathbf{X}) \sim \text{GRDPG}(n, d_+, d_-) $ if for any $ i, j \in \{ 1, \cdots, n \} $
	\begin{equation}
	\label{eq:grdpg}
	    \begin{split}
	        \mathbf{A}_{ij} & \sim \text{Bernoulli}(\mathbf{P}_{ij}), \\
	        \mathbf{P}_{ij} & = \mathbf{X}_{i}^\top \mathbf{I}_{d_+ d_-} \mathbf{X}_j.
	    \end{split}
	\end{equation}
\end{definition}

As mentioned above, the SBM, which encapsulates block structure in independent-edge networks, is a special case of the GRDPG.

\begin{definition}[$ K $-block Stochastic Blockmodel Graph~\cite{Holland1983}]
	\label{def:SBM}
	The $ K $-block stochastic blockmodel (SBM) graph is an independent-edge random graph with each vertex belonging to one of $ K $ blocks. It can be parameterized by a block connectivity probability matrix $ \mathbf{B} \in [0, 1]^{K \times K} $ and a nonnegative vector of block assignment probabilities $ \bm{\pi} \in [0, 1]^K $ summing to unity. Let $ \mathbf{A} \in \left\{0, 1 \right\}^{n \times n} $ be an adjacency matrix and $ \bm{\tau} \in \left\{1, \cdots, K \right\}^n $ be a vector of block assignments with $ \tau_i = k $ if vertex $ i $ is in block $ k $ (occurring with probability $ \pi_k $). We say $ (\mathbf{A}, \bm{\tau}) \sim \text{SBM}(n, \mathbf{B}, \bm{\pi}) $ if for any $ i, j \in \{ 1, \cdots, n \} $
	\begin{equation}
	    \begin{split}
	        \mathbf{A}_{ij} & \sim \text{Bernoulli}(\mathbf{P}_{ij}), \\
	        \mathbf{P}_{ij} & = \mathbf{B}_{\tau_i \tau_j}.
	    \end{split}
	\end{equation}
\end{definition}

 The SBM can be thought of as the GRDPG with locations fixed in each block. Formally, let $ (\mathbf{A}, \bm{\tau}) \sim \text{SBM}(n, \mathbf{B}, \bm{\pi}) $ as in Definition~\ref{def:SBM} where $ \mathbf{B} \in [0, 1]^{K \times K} $ with $ d_+ $ strictly positive eigenvalues and $ d_- $ strictly negative eigenvalues. To represent this SBM in the GRDPG model, we can choose $ \bm{\nu}_1, \cdots, \bm{\nu}_K \in \mathbb{R}^d $ where $ d = d_+ + d_- $ such that $ \bm{\nu}_k^\top \mathbf{I}_{d_+ d_-} \bm{\nu}_\ell = \mathbf{B}_{k \ell} $ for all $ k, \ell \in \{ 1, \cdots, K \} $. For example, we can take $ \bm{\nu} = \mathbf{U} |\boldsymbol{\Lambda}|^{1/2} $ where $ \mathbf{B} = \mathbf{U} \boldsymbol{\Lambda} \mathbf{U}^\top $ is the eigendecomposition of $ \mathbf{B} $ after re-ordering. Then we have the latent position of vertex $ i $ as $ \mathbf{X}_i = \bm{\nu}_k $ if $ \tau_i = k $. 

\begin{example}[Two-block Rank One Model]
	\label{example:2br1}
	As an illustration, consider the prototypical two-block SBM with rank one block connectivity probability matrix $ \mathbf{B} $ where $ \mathbf{B}_{11} = p^2, \mathbf{B}_{22} = q^2, \mathbf{B}_{12} = \mathbf{B}_{21} = pq  $ with $ 0 < p < q < 1 $. Let $ \mathbf{X}_i $ be the latent position of vertex $ i $ where $ \mathbf{X}_i = \bm{\nu}_1 = p $ if $ \tau_i = 1 $ and $ \mathbf{X}_i = \bm{\nu}_2 = q $ if $ \tau_i = 2 $. Then we can represent this SBM in the GRDPG model with latent positions $ \bm{\nu} = \begin{bmatrix} p & q \end{bmatrix}^\top $ as
	\begin{equation}
	\label{eq:B}
	\mathbf{B} = \bm{\nu} \bm{\nu}^\top 
	=
	\begin{bmatrix}
	p^2 & pq \\
	pq & q^2
	\end{bmatrix}.
	\end{equation}
\end{example}

Since our goal is to examine the impact of covariates on network inference, we next extend the  GRDPG to permit vertex covariates, as follows.

\begin{definition}[GRDPG with Vertex Covariates~\cite{Mele2019}]
	\label{def:GRDPGwithCov}
	Consider GRDPG as in Definition~\ref{def:GRDPG}. Let $ \mathbf{Z} $ denote the observed vertex covariate and $ \beta $ denote the effect of the vertex covariate. Then we say $ (\mathbf{A}, \mathbf{X}, \mathbf{Z}, \bm{\beta}) \sim \text{GRDPG-Cov}(n, d_+, d_-) $ for any $ i, j \in \{ 1, \cdots, n \} $
	\begin{equation}
	    \begin{split}
	        \mathbf{A}_{ij} & \sim \text{Bernoulli}(\mathbf{P}_{ij}), \\
	        \mathbf{P}_{ij} & = \mathbf{X}_{i}^\top \mathbf{I}_{d_+ d_-} \mathbf{X}_j + \beta \bm{1} \{ \mathbf{Z}_i=\mathbf{Z}_j \}.
	    \end{split}
	\end{equation}
\end{definition}

\begin{remark}
	\label{remark:GRDPGwithCov}
	In the case of an SBM, we have 
	\begin{equation}
	    \mathbf{P}_{ij} = \mathbf{B}_{\tau_i \tau_j} + \beta \bm{1} \{ \mathbf{Z}_i=\mathbf{Z}_j \}.
	\end{equation}
\end{remark}

\begin{example}[Two-block Rank One Model with One Binary Covariate]
	\label{example:2br1C}
	As an illustration, consider the rank one matrix $ \mathbf{B} $ in Eq.~\eqref{eq:B} and the SBM model in Remark~\ref{remark:GRDPGwithCov}. Let $ \mathbf{Z} \in \{ 1, 2 \}^{n} $ denote the observed binary covariate. Assume $ 0 < \beta < 1 $ with $ p^2+\beta, q^2+\beta, pq+\beta \in [0,1] $. Then we have the block connectivity probability matrix with the vertex covariate effect as
	\begin{equation}
	\label{eq:BZ}
	\mathbf{B_{Z}} = 
	\begin{bmatrix}
	p^2+\beta & p^2 & pq+\beta & pq \\
	p^2 & p^2+\beta & pq & pq+\beta \\
	pq+\beta & pq & q^2+\beta & q^2 \\
	pq & pq+\beta & q^2 & q^2+\beta
	\end{bmatrix}.
	\end{equation}
\end{example}

\begin{example}[Two-block Homogeneous Model with One Binary Covariate]
	\label{example:2bhC}
	As a second illustration, consider the rank two matrix $ \mathbf{B} $ where $ \mathbf{B}_{11} = \mathbf{B}_{22} = a, \mathbf{B}_{12} = \mathbf{B}_{21} = b $ with $ 0 < b < a < 1 $. Assume $ 0 < \beta < 1 $ with $ a+\beta, b+\beta \in [0,1] $. We then have the block connectivity probability matrix with the vertex covariate effect as
	\begin{equation}
	\label{eq:BZab}
	\mathbf{B_{Z}} = 
	\begin{bmatrix}
	a+\beta & a & b+\beta & b \\
	a & a+\beta & b & b+\beta \\
	b+\beta & b & a+\beta & a \\
	b & b+\beta & a & a+\beta
	\end{bmatrix}.
	\end{equation}
\end{example}

\begin{remark}
	\label{remark:kbh}
	The SBMs parameterized by $ \mathbf{B} $ in Example~\ref{example:2bhC} lead to the notion of the homogeneous model~\cite{Abbe2018,Cape2019}. For $K$-block homogeneous model, we have $ \mathbf{B}_{k \ell} = a $ for $ k = \ell $ and $ \mathbf{B}_{k \ell} = b $ for $ k \neq \ell $.
\end{remark}

In Examples 2 and 3, an induced two-block SBM becomes a four-block SBM via the effect of a binary vertex covariate. The goal is to cluster each vertex into one of the two induced blocks after accounting for the vertex covariate effect. To this end, we need to recover the latent positions of the underlying GRDPG, using the adjacency spectral embedding.

\begin{definition}[Adjacency Spectral Embedding]
	Let $ \mathbf{A} \in \{0, 1 \}^{n \times n} $ be an adjacency matrix with eigendecomposition $ \mathbf{A} = \mathbf{U} \boldsymbol{\Lambda} \mathbf{U}^\top $. Given the embedding dimension $ d < n $, the adjacency spectral embedding (ASE) of $ \mathbf{A} $ into $ \mathbb{R}^d $ is the $ n \times d $ matrix $ \mathbf{\widehat{X}} = \mathbf{\widehat{U}}_d |\boldsymbol{\widehat{\Lambda}}_d|^{1/2} $ where $ \boldsymbol{\widehat{\Lambda}}_d $ is a diagonal matrix with the $ d $ largest eigenvalues in magnitudes and $ \mathbf{\widehat{U}}_d $ contains the associated eigenvectors. Here hat notation suggests these terms estimate the eigenvectors and eigenvalues of the matrix $ \mathbf{P} $ as in Eq.~\eqref{eq:grdpg}.
\end{definition}

\begin{remark}
	\label{remark:dhat}
	There are different methods for choosing the embedding dimension~\cite{Hastie2009,Jolliffe2016}; we adopt the well-established and computationally efficient profile likelihood method~\cite{Zhu2006} to automatically identify an elbow in the scree plot to select embedding dimension $ \widehat{d} $.
\end{remark}

\section{Model-based Spectral Inference}

\label{sec:3}

We are interested in estimating the induced block assignments (clustering vertices) in a SBM with vertex covariates. To that end, we also consider algorithms for estimating the vertex covariate effect $ \beta$, which can be further used to estimate the induced block assignments. Our model-based spectral algorithms take observed adjacency matrices (and vertex covariates) as inputs and estimated block assignments for each vertex as outputs.

\begin{algorithm}
	\label{algo:1}
	\SetAlgoNoLine
	\KwIn{Adjacency matrix $ \mathbf{A} \in \{0, 1\}^{n \times n} $}
	\KwOut{Induced block assignments as $ \bm{\widehat{\tau}} $.}
	
	Estimate latent positions under the effects of both observed covariates and unobserved heterogeneity of vertices as $ \mathbf{\widehat{Y}} \in \mathbb{R}^{n \times \widehat{d}} $ using ASE of $ \mathbf{A} $ where $ \widehat{d} $ is chosen as in Remark~\ref{remark:dhat}. \\
	
	Cluster $ \mathbf{\widehat{Y}} $ using Gaussian mixture modeling (GMM) to estimate the block assignments under the effects of both observed covariates and unobserved heterogeneity of vertices
    as $ \bm{\widehat{\xi}} \in \{1, \cdots, \widehat{K} \}^{n} $ where $ \widehat{K} $ is chosen via Bayesian Information Criterion (BIC). \\
	
	Compute the estimated block connectivity probability matrix including the vertex covariate effect as
	\begin{equation*}
	\mathbf{\widehat{B}}_Z = \bm{\widehat{\mu}} \mathbf{I}_{d_+ d_-} \bm{\widehat{\mu}}^\top \in [0,1]^{\widehat{K} \times \widehat{K}},
	\end{equation*}
	where $ \bm{\widehat{\mu}} \in \mathbb{R}^{\widehat{K} \times \widehat{d}} $ is the matrix of estimated means of all clusters. \\
	
	Cluster the diagonal of $ \mathbf{\widehat{B}}_Z $ using GMM to estimate the cluster assignments of the diagonal as $ \bm{\widehat{\phi}} \in \{1, \cdots, \frac{\widehat{K}}{2} \}^{\widehat{K}} $. \\
	
	Estimate the induced block assignments as $ \bm{\widehat{\tau}} $ by $ \bm{\widehat{\tau}}_k = c $ for $ k \in \{i \; | \; \bm{\widehat{\xi}}_i = t \; \text{for} \; t \in \{j \; | \;  \bm{\widehat{\phi}}_j = c \}  \}  $ and $ c = 1, \cdots, \frac{\widehat{K}}{2} $. 
	
	\caption{Estimation of induced block assignment using only the adjacency matrix}
\end{algorithm}

In Algorithm~\ref{algo:1}, the estimation of the induced block assignments, i.e., $ \bm{\widehat{\tau}} $, depends on the estimated block connectivity probability matrix $ \mathbf{\widehat{B}}_Z $ (see Step~4 of Algorithm~\ref{algo:1} for details). 
This suggests that we may not obtain an accurate estimate of the induced block assignments if the diagonal of $ \mathbf{\widehat{B}}_Z $ does not contain enough information to distinguish the induced block structure. To address this uncertainty, we consider a modified algorithm that uses the information from vertex covariates to estimate the induced block assignments along with vertex covariate effect $ \beta $.

\begin{algorithm}
	\label{algo:2}
	\SetAlgoNoLine
	\KwIn{Adjacency matrix $ \mathbf{A} \in \{0, 1\}^{n \times n} $; observed vertex covariates $ \mathbf{Z} \in \{1, 2 \}^{n} $}
	\KwOut{Estimated vertex covariate effect as $ \widehat{\beta} $; induced block assignments as $ \bm{\widetilde{\tau}} $.}
	
	Steps 1 - 4 in Algorithm~\ref{algo:1}. \\
	
	Estimate the vertex covariate effect as $ \widehat{\beta} $ using one of the following procedures~\cite{Mele2019}. \linebreak
	  (a) Assign the block covariates as $ \mathbf{Z}_B \in \{-1, 1 \}^{\widehat{K}} $ for each block using the mode, i.e.,
	  \begin{equation*}
	  \mathbf{Z}_{B, k} = 
	  \begin{cases}
	  -1 & \text{if} \;\; n_{-1, k} \geq n_{1, k}, \\[1em]
	  1 & \text{if} \;\; n_{-1, k} < n_{1, k},
	  \end{cases}
	  \end{equation*}
	  where 
	  \begin{equation*}
	  n_{z, k}  = \sum_{i: \bm{\widehat{\xi}}_i = k} \bm{1} { \{ \mathbf{Z}_i = z \} }.
	  \end{equation*}
	  Construct pair set $ S = \{(k \ell, k \ell^\prime), k, \ell, \ell^\prime \in \{1, \cdots, \widehat{K} \} \; | \; \bm{\widehat{\phi}}_{\ell} = \bm{\widehat{\phi}}_{\ell^\prime}, \mathbf{Z}_{B, k} = \mathbf{Z}_{B, \ell}, \mathbf{Z}_{B, k} \neq \mathbf{Z}_{B, \ell^\prime} \} $. Estimate the vertex covariate effect as
	  \begin{equation*}
	  \label{eq:betaSA}
	  \widehat{\beta}_{\text{SA}} = \frac{1}{\lvert S \rvert} \sum_{(k \ell, k \ell^\prime) \in S} \mathbf{\widehat{B}}_{Z, k\ell} - \mathbf{\widehat{B}}_{Z, k \ell^\prime}.
	  \end{equation*} \linebreak
	  (b) Compute the probability that two entries from $ \mathbf{\widehat{B}}_Z $ form a pair as
	  \begin{equation*}
	  p_{k \ell, k \ell^\prime} = \frac{n_{-1, k} n_{-1, \ell} n_{1, \ell^\prime} + n_{1, k} n_{1, \ell} n_{-1, \ell^\prime} }{n_{k} n_{\ell} n_{\ell^\prime}},
	  \end{equation*}
	  where
	  \begin{equation*}
	  n_k = \sum_{i=1}^{n} \bm{1} { \{ \bm{\widehat{\xi}}_i = k \} }.
	  \end{equation*}
	  Construct pair set $ W = \{ (\ell, \ell^\prime), \ell, \ell^\prime \in \{1, \cdots, \widehat{K} \} \; | \; \bm{\widehat{\phi}}_{\ell} = \bm{\widehat{\phi}}_{\ell^\prime} \} $. Estimate the vertex covariate effect as
	  \begin{equation*}
	  \label{eq:betaWA}
	  \widehat{\beta}_{\text{WA}} = \frac{1}{\widehat{K} \lvert W \rvert} \sum_{k=1}^{\widehat{K}} \sum_{(\ell, \ell^\prime) \in W} p_{k \ell, k \ell^\prime} \left(\mathbf{\widehat{B}}_{Z, k\ell} - \mathbf{\widehat{B}}_{Z, k \ell^\prime} \right).
	  \end{equation*} \\
	  
	 Account for the vertex covariate effect by
	 \begin{equation*}
	 \mathbf{\widetilde{A}}_{ij} = \mathbf{A}_{ij} - \widehat{\beta} \bm{1} { \{ \mathbf{Z}_i=\mathbf{Z}_j \} },
	 \end{equation*}
	 where $ \widehat{\beta} $ is either $ \widehat{\beta}_{\text{SA}} $ or $ \widehat{\beta}_{\text{WA}} $. \\
	 
	 Estimate latent positions after accounting for the vertex covariate effect as $ \mathbf{\widetilde{Y}} \in \mathbb{R}^{n \times \widetilde{d}} $ using ASE of $ \mathbf{\widetilde{A}} $ where $ \widetilde{d} $ is chosen as in Remark~\ref{remark:dhat}.  \\
	 
	 Cluster $ \mathbf{\widetilde{Y}} $ using GMM to estimate the induced block assignments as $ \bm{\widetilde{\tau}} \in \{1, \cdots, \frac{\widehat{K}}{2} \}^{n} $.
	
	\caption{Estimation of induced block assignment incorporating both the adjacency matrix and the vertex covariates}
\end{algorithm}

As an illustration of estimating $ \beta $ (Step 2 in Algorithm~\ref{algo:2}), consider the block connectivity probability matrix $ \mathbf{B}_Z $ as in Eq.~\eqref{eq:BZab}. To get $ \beta $, we can 
take the difference between two specific entries of $ \mathbf{B}_Z $. For example, 
\begin{equation}
\begin{split}
\mathbf{B}_{Z, 11} - \mathbf{B}_{Z, 12} & = (a+\beta) - a = \beta, \\
\mathbf{B}_{Z, 13} - \mathbf{B}_{Z, 14} & = (b+\beta) - b = \beta.
\end{split}
\end{equation}

We can then obtain $ \widehat{\beta} $ by subtracting two specific entries of $ \mathbf{\widehat{B}}_Z $. 
However, the ASE and GMM under GRDPG model can lead to the re-ordering of $ \mathbf{\widehat{B}}_Z $. Thus we need to identify pairs first so that we subtract the correct entries. Two alternative ways to achieve this are described in Step 2(a) and 2(b).

In Step 2(a), we find pairs in $ \mathbf{\widehat{B}}_Z $ by first assigning each block common covariates using the mode. However, it is possible that we can not find any pairs using this approach, especially in the unbalanced case where the size of each block is different and/or the distribution of the vertex covariate is different. For example, one block size is much larger than the others and/or vertex covariates are all the same within one block. 

In Step 2(b), instead of first finding pairs using the mode, we only compute the probability that two entries of $ \mathbf{\widehat{B}}_Z $ form a pair. This will make the estimation more robust to extreme cases or special structure by giving different weights to pairs~\cite{Mele2019}.

\section{Spectral Inference Performance}

\label{sec:4}

\subsection{Chernoff Ratio}

\subsubsection{Main idea}

We employ Chernoff information to compare the performance of Algorithms~\ref{algo:1} and~\ref{algo:2} for estimating the induced block assignments in SBMs with vertex covariates. There are other metrics for comparing spectral inference performance such as within-class covariance. The advantages of Chernoff information are that it is independent of the clustering procedure, i.e., it can be derived no matter which clustering methods are used, and it is intrinsically related to the Bayes risk~\cite{Athreya2017,Karrer2011,Tang2018}. In short, there will be a quantity associated with each algorithm, say $ \rho_1^* $ and $ \rho_2^* $ are associated with the Algorithms~\ref{algo:1} and~\ref{algo:2} respectively. The comparison is based on the ratio $ \rho^* = \rho_1^* / \rho_2^* $. If $ \rho^* > 1 $, then Algorithm~\ref{algo:1} is preferred, otherwise Algorithm~\ref{algo:2} is preferred. The following sections provide the mathematical details of Chernoff information and derive $ \rho^* $ for certain model of interest.

\subsubsection{Mathematical details}

Let $ F_1 $ and $ F_2 $ be two continuous multivariate distributions on $ \mathbb{R}^d $ with density functions $ f_1 $ and $ f_2 $. The Chernoff information~\cite{Chernoff1952,Chernoff1956} is defined as
\begin{equation}
\begin{split}
C(F_1, F_2) & = - \log \left[\inf_{t \in (0,1)} \int_{\mathbb{R}^d} f_1^t(\mathbf{x}) f_2^{1-t}(\mathbf{x}) d\mathbf{x} \right] \\
& = \sup_{t \in (0, 1)} \left[- \log \int_{\mathbb{R}^d} f_1^t(\mathbf{x}) f_2^{1-t}(\mathbf{x}) d\mathbf{x} \right].
\end{split}
\end{equation}

Consider the special case where we take $ F_1 = \mathcal{N}(\bm{\mu}_1, \bm{\Sigma}_1) $ and $ F_2 = \mathcal{N}(\bm{\mu}_2, \bm{\Sigma}_2) $; then the corresponding Chernoff information is
\begin{equation}
\begin{split}
C(F_1, F_2) & = \sup_{t \in (0, 1)} \left[ \frac{1}{2} t (1-t) (\bm{\mu}_1 - \bm{\mu}_2)^\top \bm{\Sigma}_t^{-1} (\bm{\mu}_1 - \bm{\mu}_2) \right. \\
& \hspace{4.5em} \left. + \frac{1}{2} \log \frac{\lvert \bm{\Sigma}_t \rvert}{\lvert \bm{\Sigma}_1 \rvert^t \lvert \bm{\Sigma}_2 \rvert^{1-t}} \right],
\end{split}
\end{equation}

where $ \bm{\Sigma}_t = t \bm{\Sigma}_1 + (1-t) \bm{\Sigma}_2 $. For a given embedding method such as ASE in Algorithms~\ref{algo:1} and~\ref{algo:2}, comparison via Chernoff information is based on the statistical information between the limiting distributions of the blocks and smaller statistical information implies less information to discriminate between different blocks of the SBM. To that end, we also review the limiting results of ASE for SBM, essential for investigating Chernoff information.

\begin{theorem}[CLT of ASE for SBM~\cite{Rubin-Delanchy2017}]
	\label{thm:CLT-ASE-SBM}
	Let $ (\mathbf{A}^{(n)}, \mathbf{X}^{(n)}) \sim \text{GRDPG}(n, d_+, d_-) $ be a sequence of adjacency matrices and associated latent positions of a $ d $-dimensional GRDPG as in Definition~\ref{def:GRDPG} from an inner product distribution $ F $ where $ F $ is a mixture of $ K $ point masses in $ \mathbb{R}^d $, i.e.,
	\begin{equation}
	F = \sum_{k=1}^{K} \pi_k \delta_{\bm{\nu}_k} \quad \text{with} \quad \forall k, \; \pi_k > 0 \;\; \text{and} \;\; \sum_{k=1}^{K} \pi_k = 1,
	\end{equation}
	where $ \delta_{\bm{\nu}_k} $ is the Dirac delta measure at $ \bm{\nu}_k $. Let $ \Phi(\mathbf{z}, \bm{\Sigma}) $ denote the cumulative distribution function (CDF) of a multivariate Gaussian distribution with mean $ \bm{0} $ and covariance matrix $ \bm{\Sigma} $, evaluated at $ \mathbf{z} \in \mathbb{R}^d $. Let $ \mathbf{\widehat{X}}^{(n)} $ be the ASE of $ \mathbf{A}^{(n)} $ with $ \mathbf{\widehat{X}}^{(n)}_i $ as the $ i $-th row (same for $ \mathbf{X}^{(n)}_i $). Then there exists a sequence of matrices $ \mathbf{M}_n \in \mathbb{R}^{d \times d} $ satisfying $ \mathbf{M}_n \mathbf{I}_{d_+ d_-} \mathbf{M}_n^\top = \mathbf{I}_{d_+ d_-} $ such that for all $ \mathbf{z} \in \mathbb{R}^d $ and fixed index i,
	\begin{equation}
	\mathbb{P} \left\{ \sqrt{n} \left(\mathbf{M}_n \mathbf{\widehat{X}}^{(n)}_i - \mathbf{X}^{(n)}_i \right) \leq \mathbf{z} \; \big| \; \mathbf{X}^{(n)}_i = \bm{\nu}_k  \right\} \to \Phi(\mathbf{z}, \bm{\Sigma}_k),
	\end{equation}
	where for $ \bm{\nu} \sim F $
	\begin{equation}
	\label{eq:Sigmax}
	\begin{split}
	\bm{\Delta} & = \mathbb{E} \left[ \bm{\nu} \bm{\nu}^\top \right], \\
	\mathbf{\Gamma}_k & = \mathbb{E} \left[ \left(\bm{\nu}_k^\top \mathbf{I}_{d_+ d_-} \bm{\nu} \right) \left(1-\bm{\nu}_k^\top \mathbf{I}_{d_+ d_-} \bm{\nu} \right) \bm{\nu} \bm{\nu}^\top \right], \\
	\bm{\Sigma}_k & = \mathbf{I}_{d_+ d_-} \bm{\Delta}^{-1} \mathbf{\Gamma}_k \bm{\Delta}^{-1} \mathbf{I}_{d_+ d_-}.
	\end{split}
	\end{equation}
\end{theorem}

\begin{remark}
	\label{remark:CLT-ASE-SBM}
	If the adjacency matrix $ \mathbf{A} $ is sampled from an SBM parameterized by the block connectivity probability matrix $ \mathbf{B} $ in Eq.~\eqref{eq:B} and block assignment probabilities $ \bm{\pi} = (\pi_1, \pi_2) $ with $ \pi_1 + \pi_2 = 1 $, then as a special case for Theorem~\ref{thm:CLT-ASE-SBM}~\cite{Athreya2017,Tang2018}, we have for each fixed index $ i $,
	\begin{equation}
	\begin{split}
	\sqrt{n} \left(\widehat{X}_i - p \right) & \xrightarrow{d} \mathcal{N} \left(0, \sigma_p^2 \right) \qquad \text{if} \;\; X_i = p, \\
	\sqrt{n} \left(\widehat{X}_i - q \right) & \xrightarrow{d} \mathcal{N} \left(0, \sigma_q^2 \right) \qquad \text{if} \;\; X_i = q.
	\end{split}
	\end{equation}
	where
	\begin{equation}
	\label{eq:sigmapq}
	\begin{split}
	\sigma_p^2 & = \frac{\pi_1 p^4 (1-p^2)+\pi_2 pq^3 (1-pq)}{[\pi_1 p^2 + \pi_2 q^2]^2}, \\
	\sigma_q^2 & = \frac{\pi_1 p^3 q (1-pq)+\pi_2 q^4 (1-q^2)}{[\pi_1 p^2 + \pi_2 q^2]^2}.
	\end{split}
	\end{equation}
\end{remark}

Now for a $ K $-block SBM, let $ \mathbf{B} \in [0, 1]^{K \times K} $ be the block connectivity probability matrix and $ \bm{\pi} \in [0, 1]^K $ be the vector of block assignment probabilities. Given an $ n $ vertex instantiation of the SBM parameterized by $ \mathbf{B} $ and $ \bm{\pi} $, for sufficiently large $ n $, the large sample optimal error rate for estimating the block assignments using ASE can be measured via Chernoff information as~\cite{Athreya2017,Tang2018}
\begin{equation}
\label{eq:rho}
\begin{split}
\rho &= \min_{k \neq l} \sup_{t \in (0, 1)} \left[ \frac{1}{2} n t (1-t) (\bm{\nu}_k - \bm{\nu}_\ell)^\top \bm{\Sigma}_{k\ell}^{-1}(t) (\bm{\nu}_k - \bm{\nu}_\ell) \right. \\
& \hspace{6.5em} \left. + \frac{1}{2} \log \frac{\lvert \bm{\Sigma}_{k \ell}(t) \rvert}{\lvert \bm{\Sigma}_k \rvert^t \lvert \bm{\Sigma}_\ell \rvert^{1-t}} \right]
\end{split},
\end{equation}
where $ \bm{\Sigma}_{k\ell}(t) = t \bm{\Sigma}_k + (1-t) \bm{\Sigma}_\ell $, $ \bm{\Sigma}_k $ and $ \bm{\Sigma}_\ell $ are defined as in Eq.~\eqref{eq:Sigmax}. Also note that as $ n \to \infty $, the logarithm term in Eq.~\eqref{eq:rho} will be dominated by the other term. Then we have the Chernoff ratio as
\begin{equation}
\label{eq:rhostar}
\begin{split}
\rho^* & = \frac{\rho_1^*}{\rho_2^*} \\
& \to \frac{\underset{k \neq \ell}{\min} \underset{{t \in (0, 1)}}{\sup}  \left[ t (1-t) (\bm{\nu}_{1, k} - \bm{\nu}_{1, \ell})^\top \bm{\Sigma}_{1, k\ell}^{-1}(t) (\bm{\nu}_{1, k} - \bm{\nu}_{1, \ell}) \right]}{\underset{k \neq \ell}{\min} \underset{{t \in (0, 1)}}{\sup} \left[ t (1-t) (\bm{\nu}_{2, k} - \bm{\nu}_{2, \ell})^\top \bm{\Sigma}_{2, k\ell}^{-1}(t) (\bm{\nu}_{2, k} - \bm{\nu}_{2, \ell}) \right]}.
\end{split}
\end{equation}

Here $ \rho_1^* $ and $ \rho_2^* $ are associated with the Algorithms~\ref{algo:1} and~\ref{algo:2} respectively. If $ \rho^* > 1 $, then Algorithm~\ref{algo:1} is preferred, otherwise Algorithm~\ref{algo:2} is preferred.

\subsection{Two-block Rank One Model with One Binary Covariate}

As an illustration of using Chernoff ratio in Eq.~\eqref{eq:rhostar} to compare the performance of Algorithms~\ref{algo:1} and~\ref{algo:2} for estimating the induced block assignments, we consider the two-block SBM with one binary covariate as in Example~\ref{example:2br1C}.

\begin{proposition}
	\label{prop:1}
	For two-block rank one model with one binary covariate as in Example~\ref{example:2br1C} with the assumption that $ n_{i} = n \pi_{i} $ and $ n_{Z,j} = n \pi_{Z,j} $ for $ i \in \{1, 2 \} $ and $ j \in \{1, 2, 3, 4 \} $ where $ \bm{\pi} = (\frac{1}{2}, \frac{1}{2}) $ and $ \bm{\pi}_Z = (\frac{1}{4}, \frac{1}{4}, \frac{1}{4}, \frac{1}{4}) $, there is no tractable closed-form for Chernoff ratio as in Eq.~\eqref{eq:rhostar} but numerical experiments can be used to obtain $ \rho_1^* $ and $ \rho_2^* $ can be derived analytically as
	\begin{equation}
	\rho_2^* = \frac{(p-q)^2 (p^2 + q^2)^2}{2 \left[ \sqrt{p^2 \phi_p + q^2 \phi_{pq}} + \sqrt{ q^2 \phi_q + p^2 \phi_{pq}} \right]^2},
	\end{equation}
	where $ \sigma_p^2, \sigma_q^2 $ are defined as in Eq.~\eqref{eq:sigmapq} and 
	\begin{equation}
	\begin{split}
	\phi_p & = p^2(1-p^2), \\
	\phi_q & = q^2(1-q^2), \\
	\phi_{pq} & = pq (1-pq).
	\end{split}
	\end{equation}
\end{proposition}

Technical details of Proposition~\ref{prop:1} can be found in the appendices. Figure~\ref{fig:1} shows the Chernoff ratio when we fix $ p = 0.3 $ and take $ q \in (0.3, 0.7), \beta \in (0.1, 0.5) $ in the two-block rank one models with one binary covariate. $ \rho^* < 1 $ for most of the region while $ \rho^* > 1 $ only when $ q $ and $ \beta $ are relatively large. Recall that the performance of Algorithm~\ref{algo:1} highly depends on the estimated block connectivity probability matrix $ \mathbf{\widehat{B}}_Z $. Large $ q $ and $ \beta $ lead to a relatively well-structured $ \mathbf{\widehat{B}}_Z $ and thus Algorithm~\ref{algo:1} can have better performance in this region.

\begin{figure}
	\centering
	\includegraphics[width=3in]{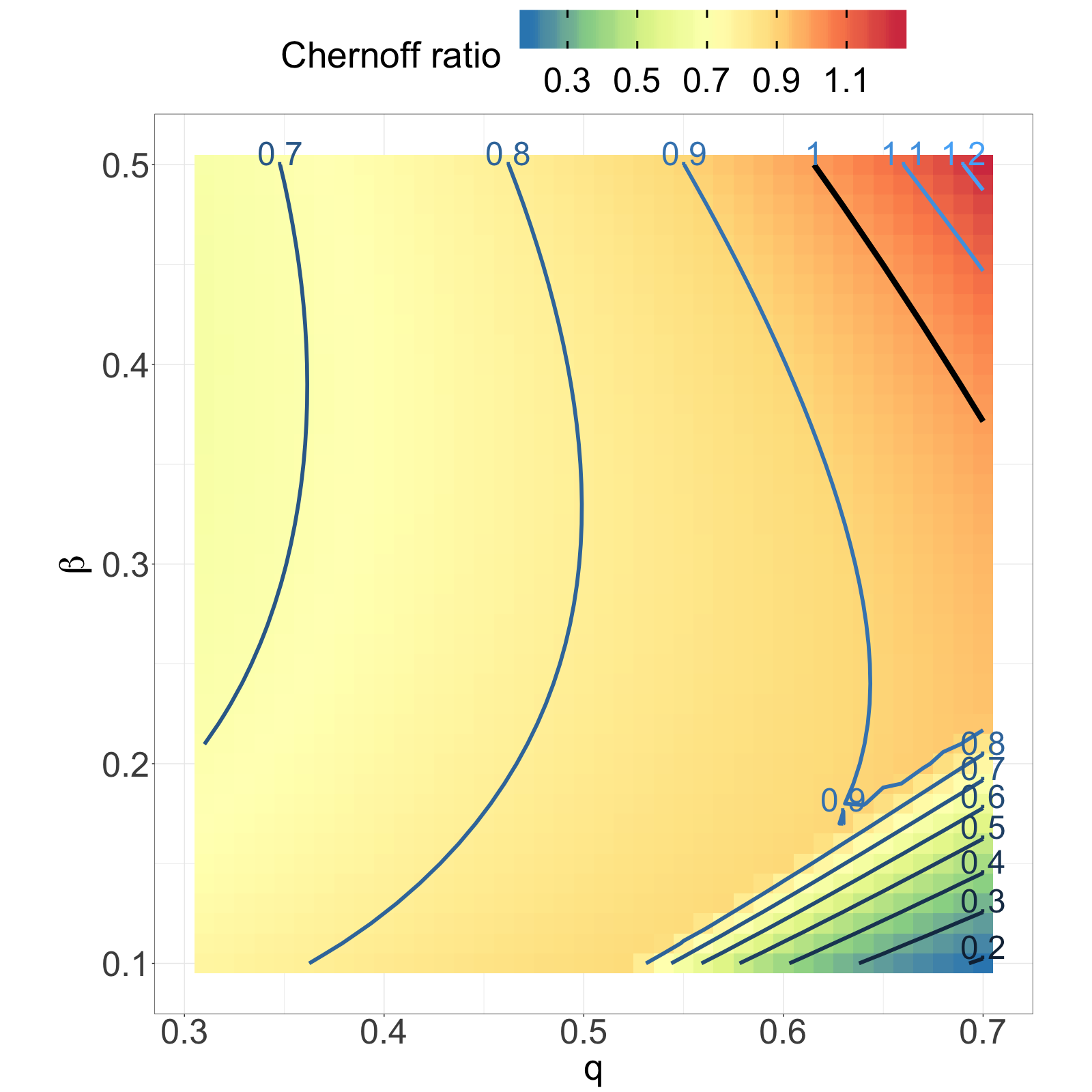}
	\caption{Chernoff ratio as in Eq.~\eqref{eq:rhostar} for two-block rank one model, $ p = 0.3, q \in (0.3, 0.7), \beta \in (0.1, 0.5), \bm{\pi} = (\frac{1}{2}, \frac{1}{2}), \bm{\pi}_Z = (\frac{1}{4}, \frac{1}{4}, \frac{1}{4}, \frac{1}{4}) $.}
	\label{fig:1}
\end{figure}

\subsection{Two-block Homogeneous Model with One Binary Covariate}

Now we consider the two-block SBM with one binary covariate parameterized by the block connectivity probability matrix $ \mathbf{B}_{Z} $ as in Eq.~\eqref{eq:BZab}. 

\begin{corollary}
	\label{corollary:rhostar2}
	For two-block homogeneous model with one binary covariate as in Example~\ref{example:2bhC} with the assumption that $ n_{i} = n \pi_{i} $ and $ n_{Z,j} = n \pi_{Z,j} $ for $ i \in \{1, 2 \} $ and $ j \in \{1, 2, 3, 4 \} $ where $ \bm{\pi} = (\frac{1}{2}, \frac{1}{2}) $ and $ \bm{\pi}_Z = (\frac{1}{4}, \frac{1}{4}, \frac{1}{4}, \frac{1}{4}) $. The Chernoff ratio as in Eq.~\eqref{eq:rhostar} can be derived analytically as
	\begin{equation}
	\label{eq:rhostar2}
	\rho^* = \frac{\rho_1^*}{\rho_2^*} \to 
	\begin{cases}
	\frac{\beta^2 (\phi_a + \phi_b)}{(a-b)^2 (\phi_a + \phi_b + \phi_{\beta})} & \text{if} \;\; \beta \leq a - b \\[1em]
	\frac{\phi_a + \phi_b}{\phi_a + \phi_b + \phi_{\beta}} & \text{if} \;\; \beta > a - b
	\end{cases},
	\end{equation}
	where
	\begin{equation}
	\label{eq:corollary1}
	\begin{split}
	\phi_a & = a(1-a), \\
	\phi_b & = b(1-b), \\
	\phi_{\beta} & = \beta (1-a-b-\beta).
	\end{split}
	\end{equation} 
\end{corollary}

Technical details of Corollary~\ref{corollary:rhostar2} can be found in appendices. Figure~\ref{fig:2} shows Chernoff ratio when we fix $ b = 0.1 $ and take $ a \in (0.1, 0.5), \beta \in (0.1, 0.5) $  in the two-block homogeneous models with one binary covariate. Again $ \rho^* < 1 $ for most of the region while $ \rho^* > 1 $ only when $ a $ and $ \beta $ are relatively large, which agrees with the general expression for Chernoff ratio as in Corollary~\ref{corollary:rhostar2}. According to Eq.~\eqref{eq:rhostar2}, we can have $ \rho^* > 1 $ only when $ \phi_{\beta} < 0 $ and this can happen only when $ a $ and $ \beta $ are relatively large. This implies that Algorithm~\ref{algo:2} is often preferred for estimating the induced block assignments.

\begin{figure}
	\centering
	\includegraphics[width=3in]{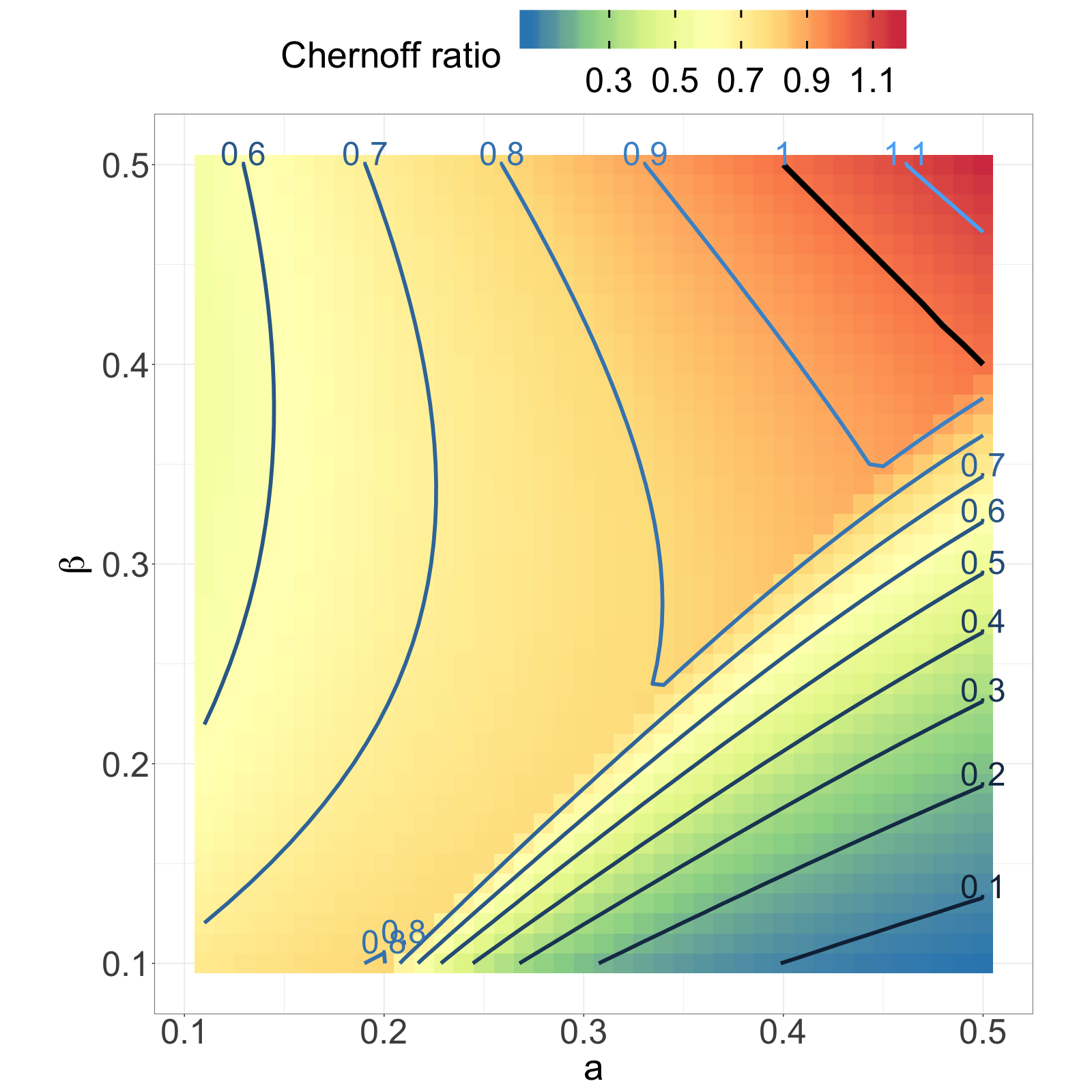}
	\caption{Chernoff ratio as in Eq.~\eqref{eq:rhostar} for two-block homogeneous models. $ b = 0.1, a \in (0.1, 0.5), \beta \in (0.1, 0.5), \bm{\pi} = (\frac{1}{2}, \frac{1}{2}), \bm{\pi}_Z = (\frac{1}{4}, \frac{1}{4}, \frac{1}{4}, \frac{1}{4}) $.}
	\label{fig:2}
\end{figure}

\subsection{$ K $-block Homogeneous Model with One Binary Covariate}

We extend the discussion from the two-block homogeneous model to the $ K $-block homogeneous model with one binary covariate. 

\begin{theorem}
	\label{thm:rhostarK}
	For the $K$-block homogeneous balanced model with one binary covariate as in Remark~\ref{remark:kbh} with the assumption that that $ n_{i} = n \pi_{i} $ and $ n_{Z,j} = n \pi_{Z,j} $ for $ i \in \{1, \cdots, K \} $ and $ j \in \{1, \cdots, 2K \} $ where $ \bm{\pi} = (\frac{1}{K}, \cdots, \frac{1}{K}) $ and $ \bm{\pi}_Z = (\frac{1}{2K}, \cdots, \frac{1}{2K}) $. The Chernoff ratio as in Eq.~\eqref{eq:rhostar} can be derived analytically as
	\begin{equation}
	\rho^* = \frac{\rho_1^*}{\rho_2^*} \to 
	\begin{cases}
	\frac{K^2 \beta^2(\phi_a+\phi_b)}{2(a-b)^2 D_4} & \text{if} \;\; \delta \leq 0 \\[1em]
	\frac{\phi_a + \phi_b}{\phi_a + \phi_b + \phi_{\beta}} & \text{if} \;\; \delta > 0
	\end{cases},
	\end{equation}
	where $ \phi_a, \phi_b, \phi_{\beta} $ are defined as in Eq.~\eqref{eq:corollary1} and
	\begin{equation}
	\begin{split}
	D_3 & = K - 2a - 2(K-1)b - K\beta, \\
	D_4 & = 2 \phi_a + 2(K-1) \phi_b + \beta D_3, \\
	\delta & = K^2 \beta^2 (\phi_a + \phi_b + \phi_{\beta}) - 2(a-b)^2 D_4.
	\end{split}
	\end{equation}
\end{theorem}

\begin{remark}
	Theorem~\ref{thm:rhostarK} generalizes Corollary~\ref{corollary:rhostar2} beyond $ K = 2 $.
\end{remark}

Technical details of Theorem~\ref{thm:rhostarK} can be found in the appendices. Figure~\ref{fig:3} shows Chernoff ratio when we fix $ b = 0.1 $ and take $ a \in (0.1, 0.5), \beta \in (0.1, 0.5) $  in the four-block homogeneous models with one binary covariate. $ \rho^* < 1 $ for most of the region while $ \rho^* > 1 $ only when $ a $ and $ \beta $ are relatively large. This implies again that Algorithm~\ref{algo:2} is often preferred for estimating the induced block assignments.

\begin{figure}
	\centering
	\includegraphics[width=3in]{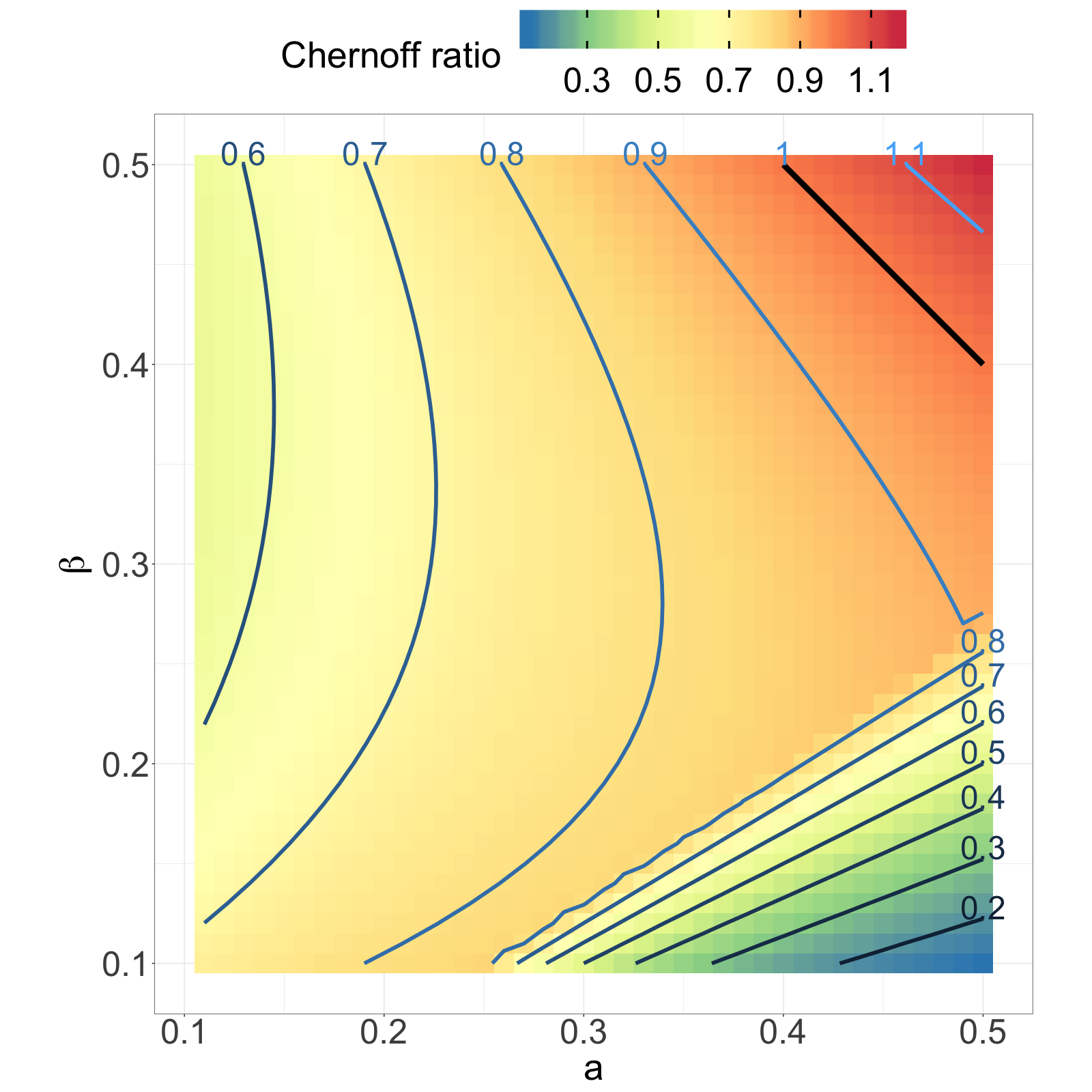}
	\caption{Chernoff ratio as in Eq.~\eqref{eq:rhostar} for four-block homogeneous models. $ b = 0.1, a \in (0.1, 0.5), \beta \in (0.1, 0.5), \bm{\pi} = (\frac{1}{4}, \frac{1}{4}, \frac{1}{4}, \frac{1}{4}), \bm{\pi}_Z = (\frac{1}{8}, \cdots, \frac{1}{8}) $.}
	\label{fig:3}
\end{figure}

\section{Simulations and Real Data Examples}

\label{sec:5}

In addition to comparing the two algorithms' performance analytically via Chernoff ratio, we also compare Algorithms~\ref{algo:1} and~\ref{algo:2} by empirical clustering results. Recall that the analytic comparison via Chernoff ratio is based on the limiting results of ASE for SBM when the number of vertices $ n \to \infty $. The comparison via empirical clustering results can measure the performance of these two algorithms for finite $ n $. The implementation of Algorithms~\ref{algo:1} and~\ref{algo:2} can be found at https://github.com/CongM/sbm-cov. Details about the experiments can be found in appendices.

\begin{figure*}
\centering
\includegraphics[width=6.5in]{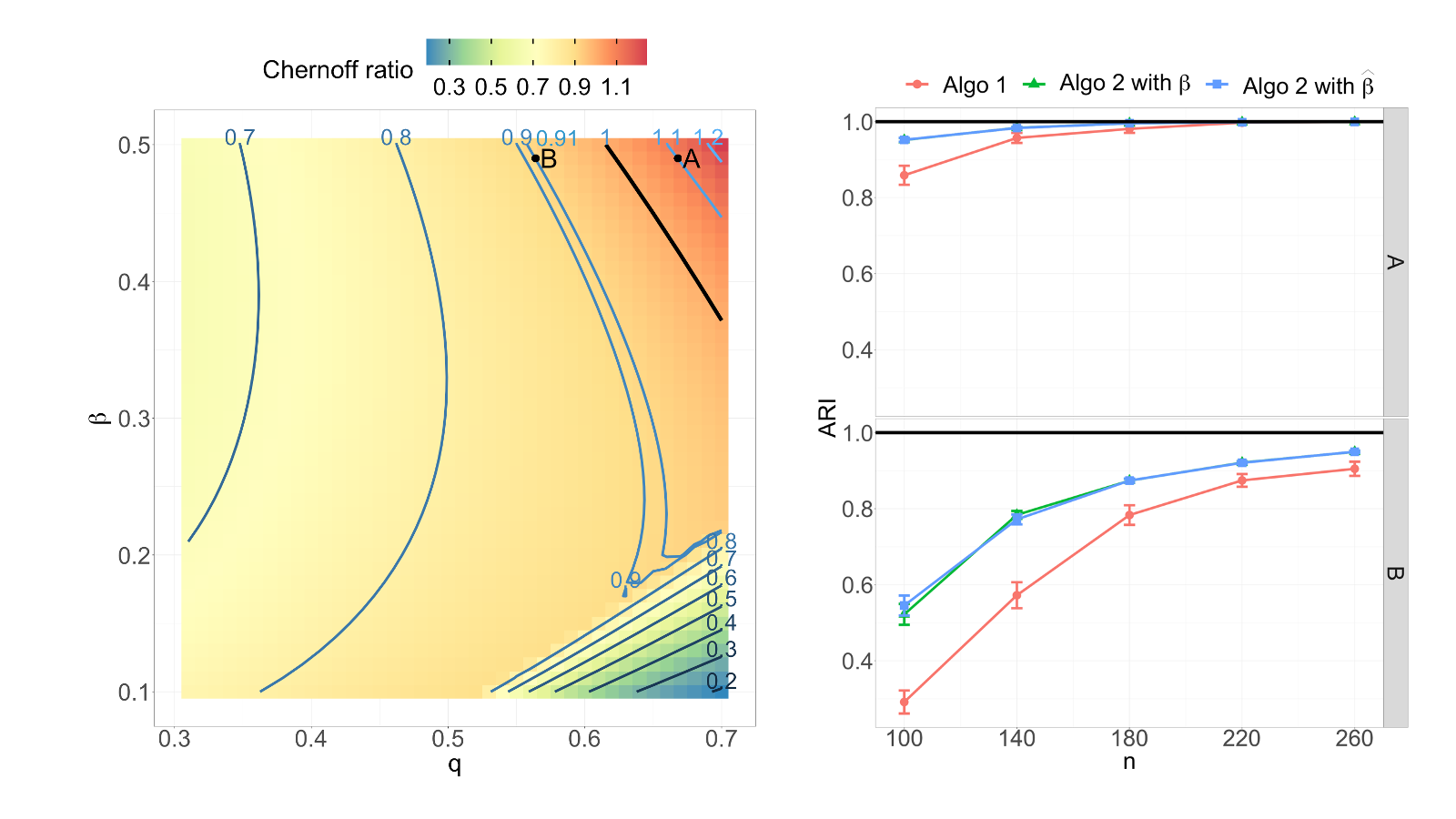}
\caption{Correspondence between Chernoff analysis and simulations.}
\label{fig:4}
\end{figure*}

As an illustration of this correspondence, we start with the setting related to ``A" ($ p = 0.3, q = 0.668, \beta = 0.49 $ with $ \rho^* = 1.1 > 1 $) and ``B" ($ p = 0.3, q = 0.564, \beta = 0.49 $ with $ \rho^* = 0.91 < 1 $) in left panel of Figure~\ref{fig:4} for two-block rank one model with one binary covariate $ \mathbf{Z} \in \{1, 2 \}^n $. We consider the balanced case where $ n_1 = n_2 = \frac{n}{2} $ and $ n_{Z, 1} = n_{Z, 2} = n_{Z, 3} = n_{Z, 4} = \frac{n}{4} $. For each $ n \in \{100, 140, 180, 220, 260 \}  $, we simulate 100 adjacency matrices with $ \frac{n}{2} $ vertices in each block and generate binary covariate with $ \frac{n}{4} $ vertices having each value of $ \mathbf{Z} $ within each block. We then apply Algorithms~\ref{algo:1} and~\ref{algo:2}  (with $ \beta $ and $ \widehat{\beta} $ in Step 3 respectively) using embedding dimension $ \widehat{d} = 3 $ to estimate the induced block assignments where adjusted Rand index (ARI)~\cite{Hubert1985} is used to measure the performance (ARI can take values from $ -1 $ to 1 where larger value indicates a better alignment of the empirical clustering and the “truth”). The upper right panel in Figure~\ref{fig:4} shows that although $ \rho^* > 1 $ and Algorithm~\ref{algo:1} should be preferred in terms of Chernoff ratio, the ARI suggests that Algorithm~\ref{algo:2} is preferred. While the Chernoff ratio is, in fact, a limit (computed as the sample size $n$ increases to infinity), the region for which $ \rho^* > 1 $ is so easy for clustering---e.g., $q-p$ is large for ``A"---that both algorithms are essentially perfect even for small $ n $. The lower right panel in Figure~\ref{fig:4} shows that Algorithm~\ref{algo:2} tends to have better performance than Algorithm~\ref{algo:1}, which agrees with the Chernoff ratio as in left figure where $ \rho^* < 1 $ and Algorithm~\ref{algo:2} is preferred. 

To further investigate the flexibility of our models and algorithms, we also discuss categorical vertex covariate.

\subsection{Two-block Rank One Model with One Five-categorical Covariate}

Consider the two-block rank one model with one five-categorical covariate $ \mathbf{Z} \in \{1, 2, 3, 4, 5 \}^n $, i.e., we have the block connectivity probability matrix $ \mathbf{B}_{Z} \in [0, 1]^{10 \times 10} $ with similar structure as in Eq.~\eqref{eq:BZ}.

\begin{figure*}
	\centering
	\subfloat[ARI as latent positions of two induced blocks move away from each other with $ \beta =0.4 $. \label{fig:5a}]{\includegraphics[width=3in]{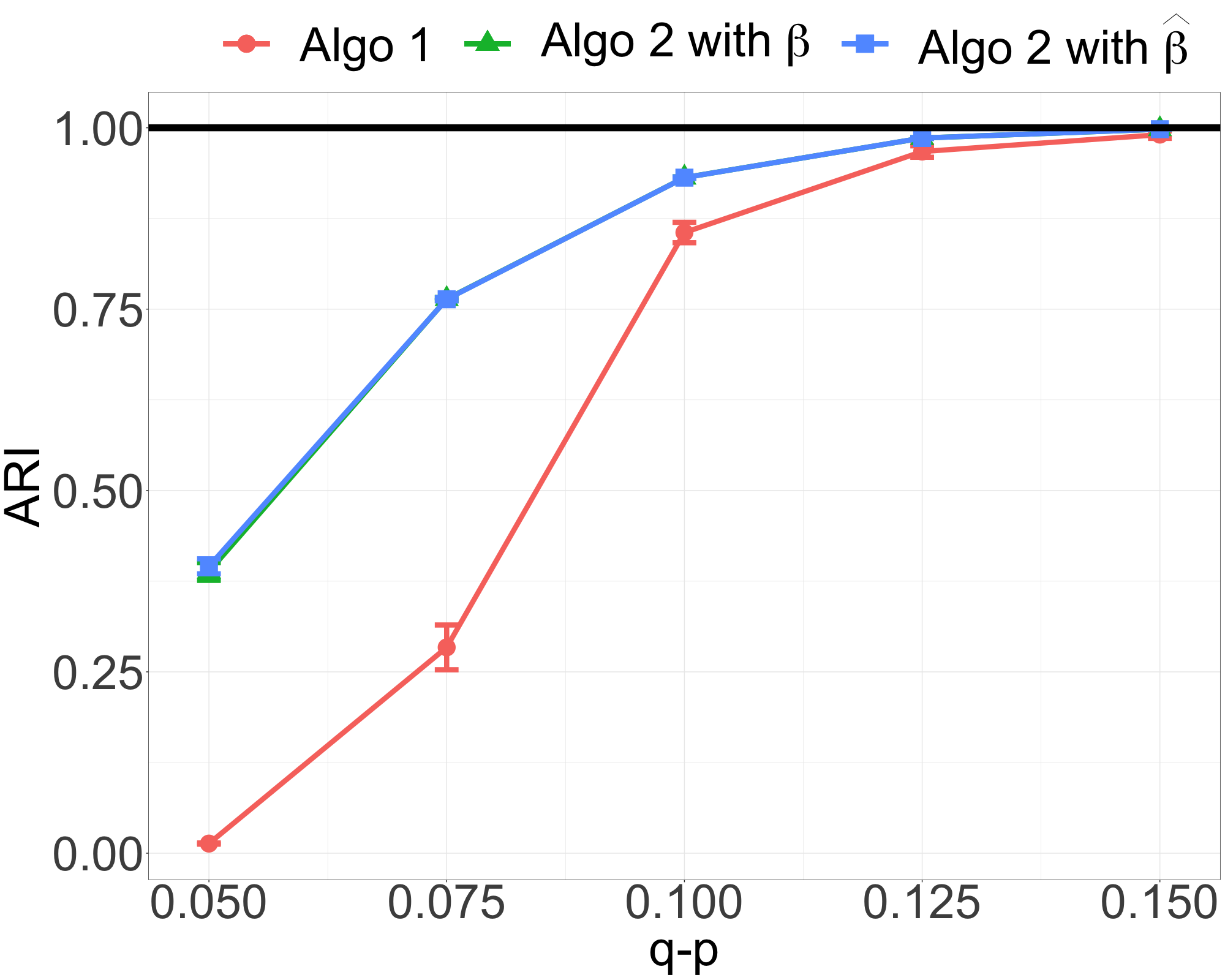}%
	}
	\hfil
	\subfloat[ARI as $ \beta $ increases with $ p = 0.3, q = 0.375 $. \label{fig:5b}]{\includegraphics[width=3in]{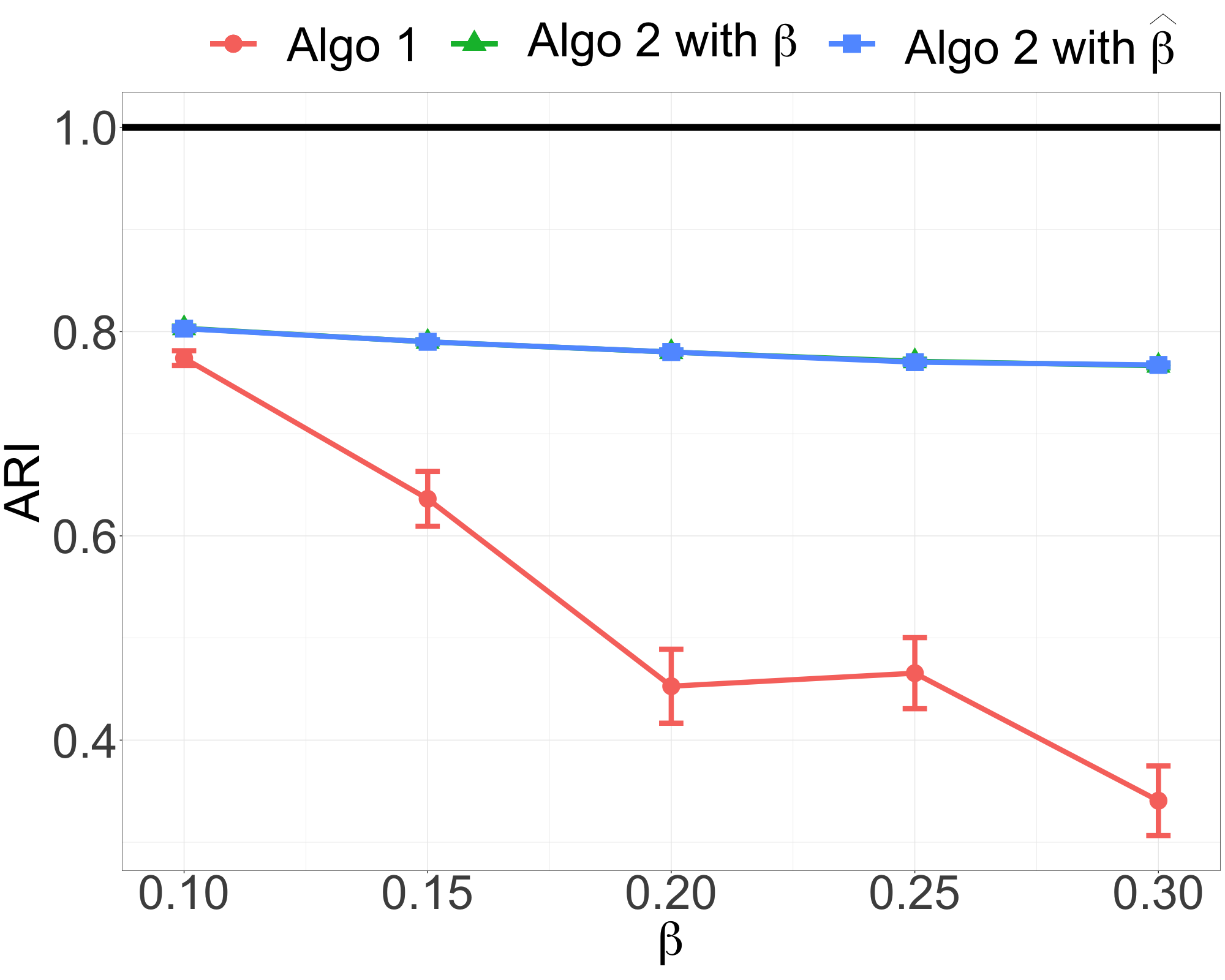}%
	}
	\caption{Simulations for two-block rank one model with one five-categorical covariate, balanced case.}
	\label{fig:5}
\end{figure*}

We first fix $ p = 0.3, \beta = 0.4 $ and consider $ q \in \{0.35, 0.375, 0.4, 0.425, 0.45 \} $. For each $ q $, we simulate 100 adjacency matrices with 1000 vertices in each block and generate five-categorical covariate with 200 vertices having each value of $ \mathbf{Z} $ within each block. We then apply Algorithms~\ref{algo:1} and~\ref{algo:2}  (with $ \beta $ and $ \widehat{\beta} $ in Step 3 respectively) using embedding dimension $ \widehat{d} = 6 $ to estimate the induced block assignments. Figure~\ref{fig:5a} shows that both algorithms estimate more accurate induced block assignments as the latent positions of two induced block move away from each other, i.e., two induced blocks tend to be more separate, and Algorithm~\ref{algo:2} can have better performance than Algorithm~\ref{algo:1}. 

Next we fix $ p = 0.3, q = 0.375 $ and consider $ \beta \in \{ 0.1, 0.15, 0.2, 0.25, 0.3 \} $. For each $ \beta $, we simulate 100 adjacency matrices with 1000 vertices in each block and generate five-categorical covariate with 200 vertices having each value of $ \mathbf{Z} $ within each block. We then apply both algorithms (with $ \beta $ and $ \widehat{\beta} $ in Step 3 of Algorithm~\ref{algo:2} respectively) using embedding dimension $ \widehat{d} = 6 $ to estimate the induced block assignments. Figure~\ref{fig:5b} shows Algorithm~\ref{algo:1} can only estimate accurate induced block assignments when $ \beta $ is relatively small while Algorithm~\ref{algo:2} can estimate accurate induced block assignments no matter $ \beta $ is small or large. Intuitively, as Algorithm~\ref{algo:1} directly estimates the induced block assignments, when $ \beta $ is relatively large, i.e., vertex covariates can affect block structure significantly, it lacks the ability to distinguish this effect. However, Algorithm~\ref{algo:2} can use additional information from vertex covariates to estimate $ \beta $, taking this effect into consideration when estimating the induced block assignments. Again, the overall performance of Algorithm~\ref{algo:2} is better than that of Algorithm~\ref{algo:1}. 

\subsection{Two-block Homogeneous Model with One Five-categorical Covariate}

We now consider the two-block homogeneous model with one five-categorical covariate $ \mathbf{Z} \in \{1, 2, 3, 4, 5 \}^n $, i.e., we have the block connectivity probability matrix $ \mathbf{B}_{Z} \in [0, 1]^{10 \times 10} $ with the similar structure as in Eq.~\eqref{eq:BZab}. Note that we can re-write $ \mathbf{B} $ like Eq.~\eqref{eq:B} as
\begin{equation}
\label{eq:Bnu}
\mathbf{B} = \bm{\nu} \bm{\nu}^\top = 
\begin{bmatrix}
a & b \\
b & a
\end{bmatrix} \;\; \text{with} \;\;
\bm{\nu} = 
\begin{bmatrix}
\sqrt{a} & 0 \\
\frac{b}{\sqrt{a}} & \sqrt{\frac{(a-b)(a+b)}{a}}
\end{bmatrix}.
\end{equation}

With these canonical latent positions, the distance between two induced blocks can be measured by
\begin{equation}
\label{eq:2ab}
\left(\sqrt{a} - \frac{b}{\sqrt{a}} \right)^2 + \left(0 - \sqrt{\frac{(a-b)(a+b)}{a}} \right)^2 = 2(a-b).
\end{equation}

\begin{figure*}
	\centering
	\subfloat[ARI as latent positions of two induced blocks move away from each other with $ \beta=0.2 $. \label{fig:6a}]{\includegraphics[width=3in]{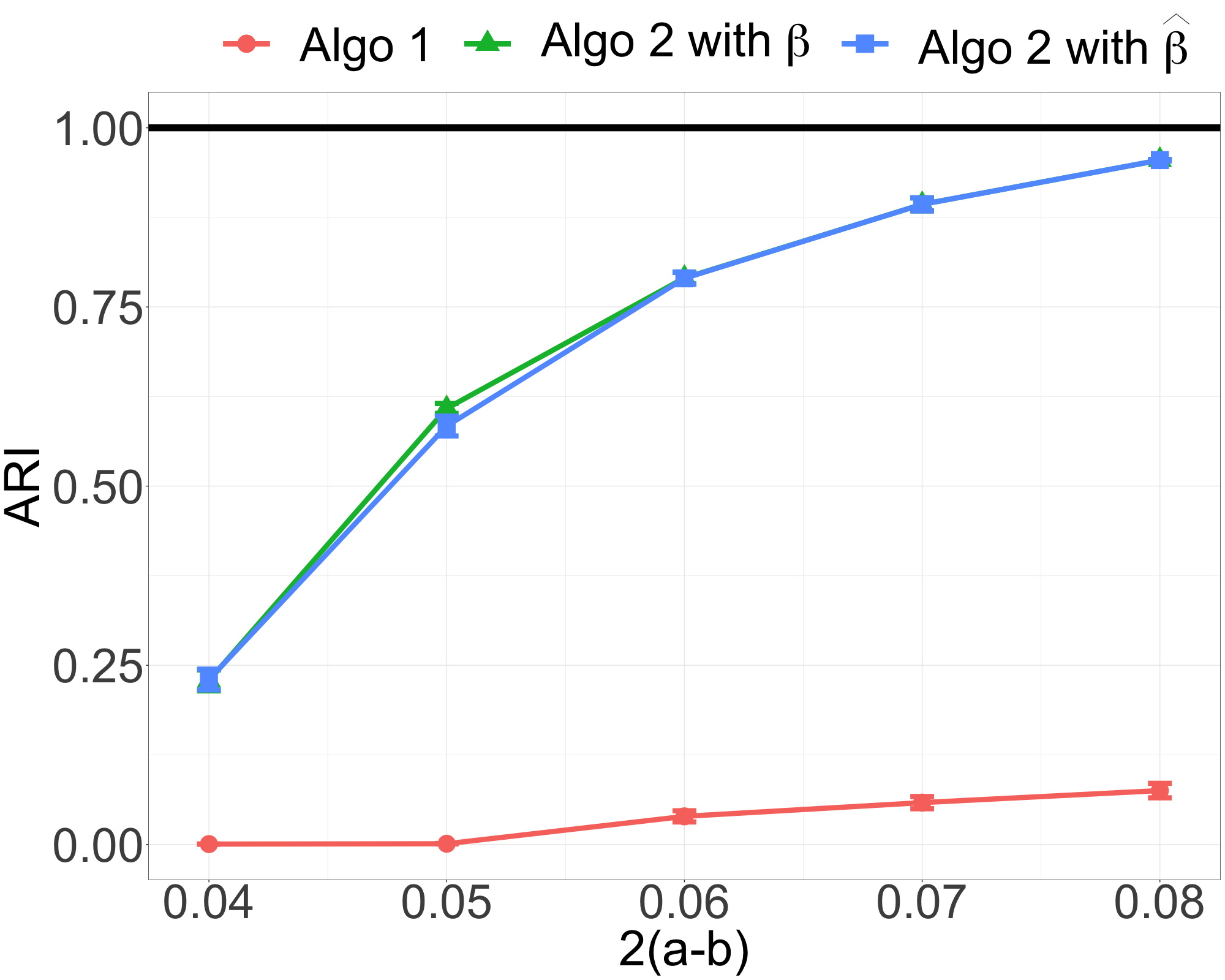}%
	}
	\hfil
	\subfloat[ARI as $ \beta $ increases with $ a = 0.135, b = 0.1 $. \label{fig:6b}]{\includegraphics[width=3in]{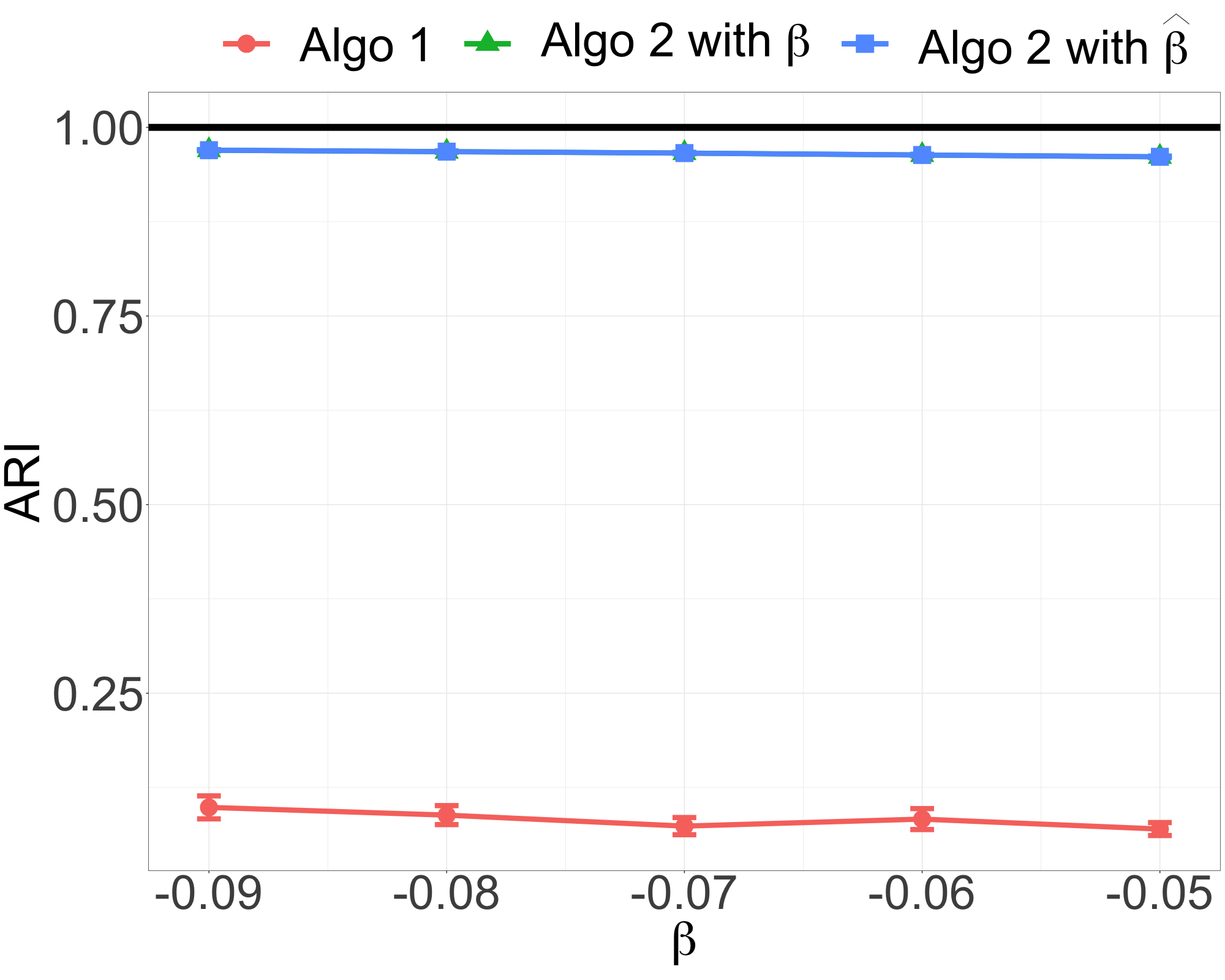}%
	}
	\caption{Simulations for two-block homogeneous model with one five-categorical covariate, balanced case.}
	\label{fig:6}
\end{figure*}

We first fix $ b = 0.1, \beta = 0.2 $ and consider $ a \in \{ 0.12, 0.125, 0.13, 0.135, 0.14 \} $. For each $ a $, we simulate 100 adjacency matrices with 1000 vertices in each block and generate five-categorical covariate with 200 vertices having each value of $ \mathbf{Z} $ within each block. We then apply both algorithms  (with $ \beta $ and $ \widehat{\beta} $ in Step 3 of Algorithm~\ref{algo:2} respectively) using embedding dimension $ \widehat{d} = 6 $ to estimate the induced block assignments. Figure~\ref{fig:6a} shows that both algorithms estimate more accurate induced block assignments as the latent positions of two induced block move away from each other, i.e., two induced blocks tend to be more separate as measured by Eq.~\eqref{eq:2ab}, and Algorithm~\ref{algo:2} can have much better performance. Recall that Algorithm~\ref{algo:1} tries to estimate the induced block assignments by clustering the diagonal of $ \mathbf{\widehat{B}}_Z $ and re-assigning the block assignments including the vertex covariate effect. For the homogeneous model, the diagonal of $ \mathbf{B}_Z $ are all the same, which can make it hard for Algorithm~\ref{algo:1} to accurately estimate the induced block assignments. But Algorithm~\ref{algo:2} is not affected by the homogeneous structure since it estiamtes the vertex covariate effect first and then estimates the induced block assignments by clustering the estimated latent positions like the canonical ones in Eq.~\eqref{eq:Bnu}. 

Next we fix $ a = 0.135, b = 0.1 $ and consider $ \beta \in \{ -0.09, -0.08, -0.07, -0.06, -0.05 \} $. For each $ \beta $, we also simulate 100 adjacency matrices with 1000 vertices in each block and generate five-categorical covariate with 200 vertices having each value of $ \mathbf{Z} $ within each block. We then apply both algorithms (with $ \beta $ and $ \widehat{\beta} $ in Step 3 of Algorithm~\ref{algo:2} respectively) using embedding dimension $ \widehat{d} = 6 $ to estimate the induced block assignments. Figure~\ref{fig:6b} shows that both algorithms are relative stable for this homogeneous model if we fix $ a $ and $ b $, due to the special structure. Still, Algorithm~\ref{algo:2} can have much better performance than Algorithm~\ref{algo:1}. 

\subsection{Connectome Data}

We consider a real data example on diffusion MRI connectome datasets~\cite{Priebe2019}. There are 114 graphs (connectomes) estimated by the NDMG pipeline~\cite{Kiar2018} in this data set where vertices represent brain sub-regions defined via spatial proximity and edges represent tensor-based fiber streamlines connecting these sub-regions. Each vertex in these graphs also has a \{Left,~Right\} hemisphere label and a \{Gray,~White\} tissue label. We treat one label as the induced block and the other one as the vertex covariate.

\begin{figure}
\centering
\includegraphics[width=3in]{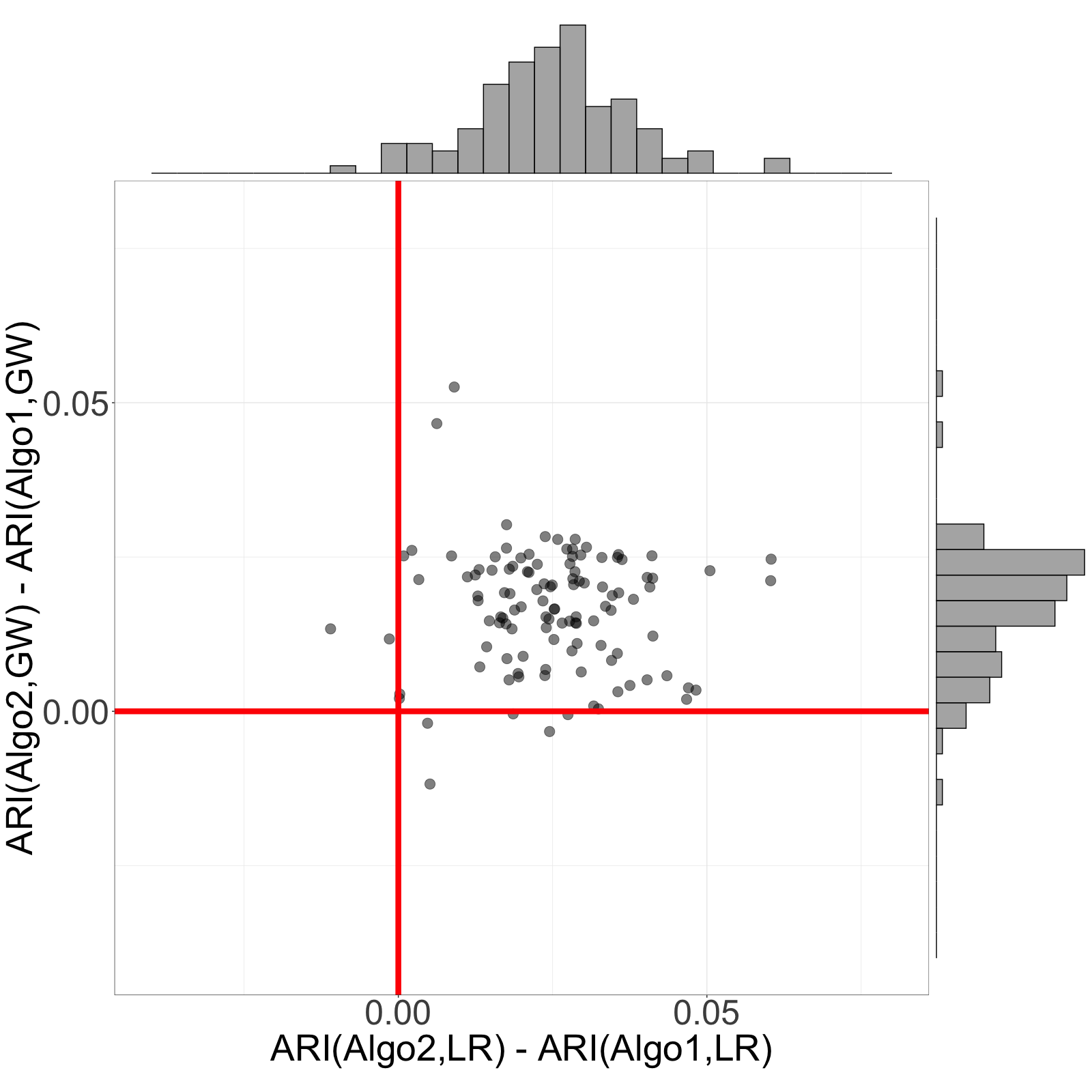}
\caption{Algorithms' comparative performance on connectome data.}
\label{fig:8}
\end{figure}

Each of the 114 connectomes (the number of vertices $ n $ varies from 23728 to 42022) is represented by a point in Figure~\ref{fig:8} with $ x = \text{ARI(Algo2, LR)} - \text{ARI(Algo1, LR)}  $ and $ y = \text{ARI(Algo2, GW)} - \text{ARI(Algo1, GW)}  $ where ARI(Algo1, LR) denotes the ARI when we apply Algorithm~\ref{algo:1} and treat \{Left,~Right\} as the induced block (with analogous notation for the rest). We see that most of the points lie in the (+,+) quadrant, indicating $  \text{ARI(Algo2, LR)} > \text{ARI(Algo1, LR)}  $ and $ \text{ARI(Algo2, GW)} > \text{ARI(Algo1, GW)} $. That is, Algorithm~\ref{algo:2} is better at estimating the induced block assignments for this real application. Note that this claim holds no matter which label is treated as the induced block. This again emphasizes the importance of distinguishing different factors that can affect block structure in graphs. Algorithm~\ref{algo:2} is able to identify particular block structure by using the observed vertex covariate information. That is, it is more likely to discover the \{Left,~Right\} structure after accounting for the effect of \{Gray,~White\} label and more likely to discover the \{Gray,~White\} structure after accounting for the effect of \{Left,~Right\} label. 

\subsection{Social Network Data}

We also utilize three social network datasets to compare our methods with several existing methods that also incorporate vertex covariates and can be scaled to deal with relatively large networks. Specifically, we compare with spectral clustering with adjacency matrix only (SCA) and covariates only (SCC)~\cite{Von2007}, pairwise covariates-adjusted stochastic blockmodel via maximum likelihood estimation (PCABM.MLE) and spectral clustering with adjustment (PCABM.SCWA)~\cite{Huang2018}, covariate-assisted spectral clustering (CASC)~\cite{Binkiewicz2017}.

\begin{itemize}
	\item LastFM asia social network dataset~\cite{Leskovec2014,Rozemberczki2020}: there are 7624 vertices that represent LastFM users from asian countries and 27806 edges that represent mutual follower relationships. We treat the location of users, which are derived from the country field for each user, as the induced block. For the vertex covariate, we focus on the number of artists liked by users, which is discretized into four categories \{0--200,~200--400,~400--600,~600+\}. 
	\item Facebook large page-page network dataset~\cite{Leskovec2014,Rozemberczki2019}: there are 22470 vertices that represent official Facebook pages and 171002 edges that represent mutual likes. We treat four page types \{Politician,~Governmental~Organization,~Television~Show,~Company\}, which are defined by Facebook, as the induced block. For the vertex covariate, we focus on the number of descriptions created by page owners to summarize the purpose of the site, which is discretized into two categories \{0--15,~15+\}. 
	\item GitHub social network dataset~\cite{Leskovec2014,Rozemberczki2019}: there are 37700 vertices that represent GitHub developers and 289003 edges that represent mutual follower relationships. We treat two developer types \{Web,~Machine~Learning\}, which are derived from the job title of each developer, as the induced block. For the vertex covariate, we focus on the number of repositories starred by developers, which is discretized into two categories \{0--18,~18+\}. 
\end{itemize}

Table~\ref{tab:snd} summarizes the algorithms' comparative performances. Algorithm~\ref{algo:2} is better at estimating the induced block assignments for all 3 datasets. This again suggests that we can better detect the block structure after accounting for the information contained in vertex covariates with our methods.

\begin{table}
	\renewcommand{\arraystretch}{1.3} 
	\caption{Algorithms' performance on social network data via ARI} 
	\label{tab:snd}
	\centering
	\begin{tabular}{cccc}
		\hline \hline
		 & LastFM & Facebook & GitHub     \\ \hline \hline 
		SCA~\cite{Von2007} & 0.229 & 0.050 & 0.000  \\
		SCC~\cite{Von2007} & 0.012 & 0.038 & 0.001  \\
		PCABM.SCWA~\cite{Huang2018} & 0.008 & -0.002 &  0.000 \\
		PCABM.MLE~\cite{Huang2018} & 0.000 & 0.004 &  -0.002 \\
		CASC~\cite{Binkiewicz2017} & 0.020 & 0.053 &  -0.043  \\
		\hline
		Algo~1 (ours)  & 0.090 & 0.036 & 0.001   \\
		Algo~2 (ours) & \textbf{0.297} & \textbf{0.076} &  \textbf{0.013} \\ \hline \hline
	\end{tabular}
\end{table}

In real data, we may not have ground truth for the block structure. Our findings suggest that we are able to discover block structure by using observed vertex covariates, which can lead to meaningful insights in widely varying applications. That is, we can better reveal underlying block structure and thus better understand the data by accounting for the vertex covariate effect.

\section{Discussion}

\label{sec:6}

We study the problem of community detection for SBMs with vertex covariates. Specifically, we consider two model-based spectral algorithms to assess the effect of observed and unobserved vertex heterogeneity on block structure in graphs. The main difference of these two algorithms in estimating the induced block assignments is whether we estimate the vertex covariate effect using the observed covariate information. To analyze the algorithms' performance, we employ Chernoff information and derive the Chernoff ratio expression for homogeneous balanced model. We also simulate multiple adjacency matrices with varied type of covariates to compare the algorithms' performance via empirical clustering accuracy measured by ARI. In addition, we conduct real data analysis on diffusion MRI connectome datasets and social network datasets. Analytic results, simulations, and real data examples suggest that the second algorithm is often preferred: we can better estimate the induced block assignments and reveal underlying block structure by using additional information contained in vertex covariates. 
Our findings also emphasize the importance of distinguishing between observed and unobserved factors that can affect block structure in graphs.

We focus on the model specified as in Definition~\ref{def:GRDPGwithCov} and Remark~\ref{remark:GRDPGwithCov} where indicator function is used to measure the vertex covariate effect and identity function is used as the link between edge probabilities and latent positions. We also investigate the flexibility of our models and algorithms by considering categorical vertex covariates. The extension from discrete vertex covariates to continuous vertex covariates is under investigation, for instance, via latent structure models ~\cite{Athreya2018}. The indicator function is used to measure the vertex covariate effect for binary and generally categorical vertex covariates under the intuition that vertices having the same covariates are more likely to form an edge between them and different functions can be adopted for the continuous vertex covariates following the similar intuition. For example, similarity and distance functions can be chosen according to the nature of different vertex covariates to measure how they can influence graph structure. One other extension is to replace the identity link with, say, the logit link function. The idea of using Chernoff information to compare algorithms' performance can be adopted for all the above generalizations and numerical evaluations can be obtained in the absence of closed-form expressions, which in turn can reveal how graph structure will affect our algorithms and provide guidelines for real application.


%

\appendices

\section{Latent Position Geometry and Chernoff Ratio}

\label{app:A}

For convenience, we first introduce the following notions. Define for $ t \in (0,1) $, 
\begin{equation}
\label{eq:gt}
\begin{split}
g(\bm{\nu}_k, \bm{\nu}) & = \left(\bm{\nu}_k^\top \bm{\nu} \right) \left(1-\bm{\nu}_k^\top \bm{\nu} \right), \\
g_t(\bm{\nu}_k, \bm{\nu}_\ell, \bm{\nu}) & = tg(\bm{\nu}_k, \bm{\nu}) + (1-t) g(\bm{\nu}_\ell, \bm{\nu}).
\end{split}
\end{equation}

Then we can re-write $ \bm{\Sigma}_k $ and $ \bm{\Sigma}_{k\ell}(t)  $ as
\begin{equation}
\begin{split}
\bm{\Sigma}_k & = \bm{\Delta}^{-1} \mathbb{E} \left[ g(\bm{\nu}_k, \bm{\nu}) \bm{\nu} \bm{\nu}^\top \right] \bm{\Delta}^{-1}, \\
\bm{\Sigma}_{k\ell}(t) & = \bm{\Delta}^{-1} \mathbb{E} \left[ g_t(\bm{\nu}_k, \bm{\nu}_\ell, \bm{\nu}) \bm{\nu} \bm{\nu}^\top \right] \bm{\Delta}^{-1}.
\end{split}
\end{equation}

We also define for $ 1 \leq k < \ell \leq K $,
\begin{equation}
\label{eq:Ckl}
C_{k\ell} = \sup_{t \in (0, 1)} t (1-t) (\bm{\nu}_k - \bm{\nu}_\ell)^\top \bm{\Sigma}_{k\ell}^{-1}(t) (\bm{\nu}_k - \bm{\nu}_\ell).
\end{equation}

\subsection{Proof of Proposition 1}

\begin{proof}
	Via the idea of Cholesky decomposition, we can re-write $ \mathbf{B}_{Z} $ as
	\begin{equation}
	\mathbf{B}_{Z} = \bm{\nu}_Z \bm{\nu}_Z^\top =
	\begin{bmatrix}
	\bm{\nu}_1^\top \bm{\nu}_1 & \bm{\nu}_1^\top \bm{\nu}_2 & \bm{\nu}_1^\top \bm{\nu}_3 & \bm{\nu}_1^\top \bm{\nu}_4 \\
	\bm{\nu}_2^\top \bm{\nu}_1 & \bm{\nu}_2^\top \bm{\nu}_2 & \bm{\nu}_2^\top \bm{\nu}_3 & \bm{\nu}_2^\top \bm{\nu}_4 \\
	\bm{\nu}_3^\top \bm{\nu}_1 & \bm{\nu}_3^\top \bm{\nu}_2 & \bm{\nu}_3^\top \bm{\nu}_3 & \bm{\nu}_3^\top \bm{\nu}_4 \\
	\bm{\nu}_4^\top \bm{\nu}_1 & \bm{\nu}_4^\top \bm{\nu}_2 & \bm{\nu}_4^\top \bm{\nu}_3 & \bm{\nu}_4^\top \bm{\nu}_4 \\
	\end{bmatrix},
	\end{equation}
	where $ \bm{\nu}_Z = \begin{bmatrix} \bm{\nu}_1 & \bm{\nu}_2 & \bm{\nu}_3 & \bm{\nu}_4 \end{bmatrix}^\top $. Elementary calculations yield the canonical latent positions as
	\begin{equation}
	\label{eq:nuZ}
	\bm{\nu}_Z = 
	\begin{bmatrix}[1.5]
	\sqrt{p^2 + \beta} & 0 & 0 \\
	\frac{p^2}{\sqrt{p^2 + \beta}} & \sqrt{\frac{\beta (2p^2 + \beta)}{p^2 + \beta}} & 0 \\
	\frac{pq + \beta}{\sqrt{p^2 + \beta}} & \sqrt{\frac{\beta p^2 (q-p)^2}{(p^2 + \beta)(2p^2 + \beta)}} & \sqrt{\frac{\beta (q-p)^2}{(2p^2 + \beta)}} \\
	\frac{pq}{\sqrt{p^2 + \beta}} & \sqrt{\frac{\beta (p^2 + pq + \beta)^2}{(p^2 + \beta)(2p^2 + \beta)}} & \sqrt{\frac{\beta (q-p)^2}{(2p^2 + \beta)}}
	\end{bmatrix}
	.
	\end{equation}
	
	For this model, the block connectivity probability matrix $ \mathbf{B}_{Z} $ is positive semidefinite with $ \text{rank}(\mathbf{B}_{Z}) = 3 $. Then we have $  \mathbf{I}_{d_+ d_-} = \mathbf{I}_3 $ and we can omit it in our analytic derivations. With the canonical latent positions in Eq.~\eqref{eq:nuZ}, the only remaining term to derive for Chernoff ratio is $ \bm{\Sigma}_{k\ell}(t) $. 
	
	By the symmetric structure of $ \mathbf{B}_{Z} $ and the balanced assumption, we observe that $ C_{13} = C_{24}, C_{14} = C_{23} $. Thus we need only to evaluate $ C_{12}, C_{13}, C_{14}, C_{34} $. Subsequent calculations and simplification yield
	\begin{equation}
	\label{eq:Cklpq}
	\begin{split}
	C_{12} & = \frac{\beta^2}{2 [\phi_p + \phi_{pq} + \beta(1-p^2-pq-\beta)]}, \\
	C_{34} & = \frac{\beta^2}{2 [\phi_q + \phi_{pq} + \beta(1-q^2-pq-\beta)]},
	\end{split}
	\end{equation}
	where for $ 0 < p < q < 1 $
	\begin{equation}
	\label{eq:phipqbeta}
	\begin{split}
	\phi_p & = p^2(1-p^2), \\
	\phi_q & = q^2(1-q^2), \\
	\phi_{pq} & = pq (1-pq).
	\end{split}
	\end{equation}
	
	Then we have the approximate Chernoff information for Algorithm~1 as
	\begin{equation}
	\label{eq:rho1pq}
	\rho_1^* \approx \min_{k \in \{1, 3 \} , k < \ell \leq 4} C_{k\ell},
	\end{equation}
	where $ C_{k\ell} $ for $ k \in \{1, 3 \} , k < \ell \leq 4 $ are defined as in Eq.~\eqref{eq:Cklpq}. For this model, there is no tractable closed-form analytic expression for $ C_{13} $ and $ C_{14} $, so we instead obtain values  $ \rho_1^* $ by numerically solving the above optimization problem. By Remark~4 and similar calculations~\cite{Athreya2017,Tang2018}, we have the approximate Chernoff information for Algorithm~2 as
	\begin{equation}
	\label{eq:rho2pq}
	\begin{split}
	\rho_2^* & \approx \sup_{t \in (0, 1)}  t (1-t) (p-q)^2 \left[t \sigma_p^2 + (1-t) \sigma_q^2 \right]^{-1} \\
	& = \frac{(p-q)^2 (p^2 + q^2)^2}{2 \left[ \sqrt{p^2 \phi_p + q^2 \phi_{pq}} + \sqrt{ q^2 \phi_q + p^2 \phi_{pq}} \right]^2},
	\end{split}
	\end{equation}
	where $ \phi_p, \phi_q, \phi_{pq} $ are defined as in Eq.~\eqref{eq:phipqbeta} and
	\begin{equation}
	\begin{split}
	\sigma_p^2 & = \frac{\pi_1 p^4 (1-p^2)+\pi_2 pq^3 (1-pq)}{[\pi_1 p^2 + \pi_2 q^2]^2}, \\
	\sigma_q^2 & = \frac{\pi_1 p^3 q (1-pq)+\pi_2 q^4 (1-q^2)}{[\pi_1 p^2 + \pi_2 q^2]^2}.
	\end{split}
	\end{equation}
\end{proof}

\subsection{Proof of Corollary 1}

\begin{proof}
	Similarly, the idea of Cholesky decomposition and elementary calculations yield the canonical latent positions as
	\begin{equation}
	\label{eq:nuZab}
	\bm{\nu}_Z = 
	\begin{bmatrix}[1.5]
	\sqrt{a + \beta} & 0 & 0 \\
	\frac{a}{\sqrt{a + \beta}} & \sqrt{\frac{\beta (2a + \beta)}{a + \beta}} &	0 \\
	\frac{b + \beta}{\sqrt{a + \beta}} & \sqrt{\frac{\beta (b-a)^2}{(a + \beta)(2a + \beta)}} & \sqrt{\frac{2(a-b)(a+b+\beta)}{(2a + \beta)}} \\
	\frac{b}{\sqrt{a + \beta}} & \sqrt{\frac{\beta (a + b + \beta)^2}{(a + \beta)(2a + \beta)}} &
	\sqrt{\frac{2(a-b)(a+b+\beta)}{(2a + \beta)}}
	\end{bmatrix}
	.
	\end{equation}
	
	Observe that for this model, the block connectivity probability matrix $ \mathbf{B}_{Z} $ is also positive semidefinite with $ \text{rank}(\mathbf{B}_{Z}) = 3 $. Then we have $  \mathbf{I}_{d_+ d_-} = \mathbf{I}_3 $ and we can omit it in the derivations as for two-block rank one model. To evaluate the Chernoff ratio, we also investigate the $ C_{k\ell} $ as defined in Eq.~\eqref{eq:Ckl}. Similar observations suggest that $ C_{12} = C_{34}, C_{13} = C_{24}, C_{14} = C_{23} $. Thus we only need to evaluate $ C_{12}, C_{13}, C_{14} $. Subsequent calculations and simplification yield
	\begin{equation}
	\label{eq:Cklab}
	\begin{split}
	C_{12} & = \frac{\beta^2}{2 (\phi_a + \phi_b + \phi_{\beta})}, \\
	C_{13} & = \frac{(a-b)^2}{2 (\phi_a + \phi_b + \phi_{\beta})}, \\
	C_{14} & = \frac{\beta^2 N_1+(a-b)N_2}{2[D_1+(\phi_a+\phi_b)(\phi_a+\phi_b+2\phi_{\beta})]},
	\end{split}
	\end{equation}
	where for $ 0 < b < a < 1 $ and $ 0 < \beta < 1 $
	\begin{equation}
	\label{eq:phiabbeta}
	\begin{split}
	\phi_a & = a(1-a), \\
	\phi_b & = b(1-b), \\
	\phi_{\beta} & = \beta (1-a-b-\beta), \\
	N_1 & = a(1-b)+b(1-a)+ \phi_{\beta} \\
	N_2 & = ab(a-b)+\phi_a(a+\beta)-\phi_b(b+\beta) \\
	D_1 & = \beta^2(1-2a-\beta)(1-2b-\beta).
	\end{split}
	\end{equation}
	
	Then we have the approximate Chernoff information for Algorithm~1 as given by
	\begin{equation}
	\label{eq:rho1ab}
	\rho_1^* \approx \min_{\ell \in \{2, 3, 4 \}} C_{1\ell},
	\end{equation}
	where $ C_{1\ell} $ for $ \ell \in \{2, 3, 4 \} $ are defined as in Eq.~\eqref{eq:Cklab}. Also observe that
	\begin{equation}
	\begin{split}
	C_{12} - C_{14} & = \frac{-(a-b)^2 [\phi_a + \phi_b + \beta(1-a-b)]^2}{D_2}, \\
	C_{13} - C_{14} & = \frac{- \beta^2 N_1^2}{D_2},
	\end{split}
	\end{equation}
	where
	\begin{equation}
	D_2 = 2 (\phi_a + \phi_b + \phi_{\beta}) [D_1+(\phi_a+\phi_b)(\phi_a+\phi_b+2\phi_{\beta})].
	\end{equation}
	
	Then we can further simplify $ \rho_1^* $ as 
	\begin{equation}
	\rho_1^* \approx
	\begin{cases}
	\frac{\beta^2}{2 (\phi_a + \phi_b + \phi_{\beta})} & \text{if} \;\; \beta \leq a - b, \\[1em]
	\frac{(a-b)^2}{2 (\phi_a + \phi_b + \phi_{\beta})} & \text{if} \;\; \beta > a - b.
	\end{cases}
	\end{equation}
	
	By the same derivations~\cite{Cape2019}, we have the approximate Chernoff information for Algorithm~2 as
	\begin{equation}
	\rho_2^* \approx \frac{ (a-b)^2}{2 \left[a(1-a) + b(1-b) \right]} = \frac{(a-b)^2}{2 (\phi_a + \phi_b)},
	\end{equation}
	where $ \phi_a $ and $ \phi_b $ are defined as in Eq.~\eqref{eq:phiabbeta}.
\end{proof}

\subsection{Proof of Theorem 2}

\begin{proof}
	Observe that a $ K $-block SBM can become a $ 2K $-block SBM when adding a binary covariate. To analytically derive the Chernoff ratio for the $ K $-block homogeneous model with one binary covariate, we first investigate the canonical latent positions for this model via the idea of Cholesky decomposition. Specifically, let $ \mathbf{B} \in [0, 1]^{K \times K} $ denote the block connectivity probability matrix after accounting for the vertex covariate effect and $ \mathbf{B}_{Z} \in [0, 1]^{2K \times 2K} $ denote the block connectivity probability matrix including the vertex covariate effect. Here we focus on canonical latent positions for $ \mathbf{B}_{Z} $, details about the canonical latent positions for $ \mathbf{B} $ have been discussed~\cite{Cape2019}. Let $ \bm{\nu}_{Z}(K,2K) $ denote the canonical latent position matrix, then we can re-write $ \mathbf{B}_{Z} $ as 
	\begin{equation}
	\mathbf{B}_{Z} = \bm{\nu}_{Z}(K,2K) \bm{\nu}_{Z}(K,2K)^\top,
	\end{equation}
	where $ \bm{\nu}_{Z}(K,2K) = \begin{bmatrix} \bm{\nu}_1 & \cdots & \bm{\nu}_{2K} \end{bmatrix}^\top $. For $ K = 2 $ we have via the idea of Cholesky decomposition
	\begin{equation}
	\bm{\nu}_Z(2,4) = 
	\begin{bmatrix}[1.5]
	\sqrt{a + \beta} & 0 & 0 \\
	\frac{a}{\sqrt{a + \beta}} & \sqrt{\frac{\beta (2a + \beta)}{a + \beta}} &	0 \\
	\frac{b + \beta}{\sqrt{a + \beta}} & \sqrt{\frac{\beta (b-a)^2}{(a + \beta)(2a + \beta)}} & \sqrt{\frac{2(a-b)(a+b+\beta)}{(2a + \beta)}} \\
	\frac{b}{\sqrt{a + \beta}} & \sqrt{\frac{\beta (a + b + \beta)^2}{(a + \beta)(2a + \beta)}} &
	\sqrt{\frac{2(a-b)(a+b+\beta)}{(2a + \beta)}}
	\end{bmatrix}.
	\end{equation}
	
	And by induction, for $ K \geq 3 $ we have
	\begin{equation}
	\label{eq:nuZK2K}
	\begin{split}
	\bm{\nu}_Z(K,2K)_{\cdot, 1} & = 
	\begin{bmatrix}[1.5]
	\bm{\nu}_Z(K-1,2K-2)_{\cdot, 1:(K-1)}  \\
	\bm{\nu}_Z(K-1,2K-2)_{2K-3, 1:(K-1)} \\
	\bm{\nu}_Z(K-1,2K-2)_{2K-2, 1:(K-1)}
	\end{bmatrix}, \\[1em]
	\bm{\nu}_Z(K,2K)_{\cdot, 2} & = 
	\begin{bmatrix}[1.5]
	\bm{\nu}_Z(K-1,2K-2)_{\cdot, K} \\
	\kappa \bm{\nu}_Z(K-1,2K-2)_{2K-3, K} \\
	\kappa \bm{\nu}_Z(K-1,2K-2)_{2K-2, K}
	\end{bmatrix}, \\[1em]
	\bm{\nu}_Z(K,2K)_{\cdot, 3} & = 
	\begin{bmatrix}[1.5]
	\bm{0} \\
	\sqrt{\frac{(a-b)[2a+2(K-1)b+K\beta]}{2a+2(K-2)b+(K-1)\beta}} \\
	\sqrt{\frac{(a-b)[2a+2(K-1)b+K\beta]}{2a+2(K-2)b+(K-1)\beta}} 
	\end{bmatrix},
	\end{split}
	\end{equation}
	where
	\begin{equation}
	\kappa = \frac{2b+\beta}{2a+2(K-2)b+(K-1)\beta}.
	\end{equation}
	
	For this $ K $-block homogeneous model with one binary covariate, the symmetric structure of $ \mathbf{B}_Z $ yields
	\begin{equation}
	\begin{split}
	\bm{\nu}_1^\top \bm{\nu}_1 = \bm{\nu}_2^\top \bm{\nu}_2 = \cdots = \bm{\nu}_{2K}^\top \bm{\nu}_{2K} & = a + \beta, \\
	\bm{\nu}_1^\top \bm{\nu}_2 = \bm{\nu}_3^\top \bm{\nu}_4 = \cdots = \bm{\nu}_{2K-1}^\top \bm{\nu}_{2K} & = a, \\
	\bm{\nu}_1^\top \bm{\nu}_3 = \bm{\nu}_1^\top \bm{\nu}_5 = \cdots = \bm{\nu}_{2K-2}^\top \bm{\nu}_{2K} & = b + \beta, \\
	\bm{\nu}_1^\top \bm{\nu}_4 = \bm{\nu}_1^\top \bm{\nu}_6 = \cdots = \bm{\nu}_{2K-2}^\top \bm{\nu}_{2K-1} & = b.
	\end{split}
	\end{equation}
	
	Along with the balanced assumption, i.e.,~$ \bm{\pi}_Z = (\frac{1}{2K}, \cdots, \frac{1}{2K}) $, the first four rows of $ \bm{\nu}_Z(K,2K) $ are ideal for derivation as they have the fewest non-zero entries and can represent all the possible geometric structure. In other word, we can only evaluate $ C_{12}, C_{13}, C_{14} $ where $ C_{k\ell} $ is defined as in Eq.~\eqref{eq:Ckl} to derive the Chernoff ratio.
	

	For $ K $-block homogeneous model with one binary covariate, we observe that $ \mathbf{B}_{Z} $ has eigenvalue $0$ with algebraic multiplicity $ K-1 $, eigenvalue $ K \beta $ with algebraic multiplicity $1$, eigenvalue $ 2(a-b) $ with algebraic multiplicity $ K-1 $ and eigenvalue $ 2a+2(K-1)b+K\beta $ with algebraic multiplicity 1. Along with the assumption that $ 0 < b < a < 1 $ and $ 0 < \beta < 1 $, we have among non-zero eigenvalues of $ \mathbf{B}_{Z}  $
	\begin{equation}
	\begin{split}
	\lambda_{\text{max}}\left(\mathbf{B}_{Z} \right) & = 2a+2(K-1)b+K\beta, \\
	\lambda_{\text{min}}\left(\mathbf{B}_{Z} \right) & = 
	\begin{cases}
	K \beta & \text{if} \;\; \beta \leq \frac{2(a - b)}{K}, \\
	2(a - b) & \text{if} \;\; \beta > \frac{2(a - b)}{K}.
	\end{cases}
	\end{split}
	\end{equation}
	
	Thus $ \mathbf{B}_{Z} $ is positive semidefinite with $ \text{rank}(\mathbf{B}_{Z}) = K+1 $. Then we have $  \mathbf{I}_{d_+ d_-} = \mathbf{I}_{K+1} $ which has no complicating effect on the subsequent derivations. As discussed in the previous section, we only consider the first four rows of the canonical latent position matrix $ \bm{\nu_Z}(K, 2K) $ and evalute $ C_{12}, C_{13}, C_{14} $. With the definition as in Eq.~\eqref{eq:gt}, we have
	\begin{equation}
	\begin{split}
	\mathbb{E} \left[ g_{\frac{1}{2}}(\bm{\nu}_1, \bm{\nu}_2, \bm{\nu}) \bm{\nu} \bm{\nu}^\top \right] 
	& = c_0 \bm{\Delta} + c_{12} \mathbf{N}_{12} \mathbf{N}_{12}^\top, \\[1em]
	\mathbb{E} \left[ g_{\frac{1}{2}}(\bm{\nu}_1, \bm{\nu}_3, \bm{\nu}) \bm{\nu} \bm{\nu}^\top \right] 
	& = \bm{\Delta}_T + c_{13} \mathbf{N}_{13} \mathbf{N}_{13}^\top + c_{24} \mathbf{N}_{24} \mathbf{N}_{24}^\top, \\[1em]
	\mathbb{E} \left[ g_{\frac{1}{2}}(\bm{\nu}_1, \bm{\nu}_4, \bm{\nu}) \bm{\nu} \bm{\nu}^\top \right] 
	& = c_0 \bm{\Delta} + c_{14} \mathbf{N}_{14} \mathbf{N}_{14}^\top + c_{23} \mathbf{N}_{23} \mathbf{N}_{23}^\top,
	\end{split}
	\end{equation}
	where $ \bm{\Delta} \in \mathbb{R}^{(K+1) \times (K+1)} $ is defined as in Theorem~1, $ \bm{\nu}_Z \in \mathbb{R}^{2K \times (K+1)} $ is defined as in Eq.~\eqref{eq:nuZK2K} and 
	\begin{equation}
	\label{eq:appB-1}
	\begin{split}
	\phi_{a} & = a(1-a), \\ 
	\phi_{b} & = b(1-b), \\
	\phi_{b\beta} & = (b+\beta)(1-b-\beta), \\
	c_{0} & = \frac{\phi_{b} +\phi_{b\beta}}{2}, \\
	c_{12} & = \frac{(a-b)(1-a-b-\beta)}{2K}, \\
	c_{13} = c_{14} & = \frac{(a-b)(1-a-b-2\beta)}{4K} , \\
	c_{23} = c_{24} & = \frac{\phi_{a}-\phi_{b}}{4K}, \\
	c_{T} & = \frac{\beta(1-2b-\beta)}{4K}, \\
	\mathbf{N}_{k\ell} & = 
	\begin{bmatrix}
	\bm{\nu}_k & \bm{\nu}_\ell
	\end{bmatrix}
	\in \mathbb{R}^{(K+1) \times 2}, \\
	\mathbf{I}_{T} & = \text{diag}(1, -1, \cdots, 1, -1) \in \mathbb{R}^{2K \times 2K}, \\
	\bm{\Delta}_{T} & = \bm{\nu}_Z^\top \left(c_T  \mathbf{I}_{T} + \frac{c_0}{2K} \mathbf{I}_{2K} \right) \bm{\nu}_Z \in \mathbb{R}^{(K+1) \times (K+1)}.
	\end{split}
	\end{equation}
	
	With the canonical latent position matrix $ \bm{\nu}_Z(K,2K) $ as in Eq.~\eqref{eq:nuZK2K}, observe that
	\begin{equation}
	\begin{split}
	\mathbf{N}_{12}^\top \bm{\Delta}^{-1} \mathbf{N}_{12} & = 
	\begin{bmatrix}
	K+1 & K-1 \\
	K-1 & K+1
	\end{bmatrix}, \\[1em]
	\mathbf{N}_{13}^\top \bm{\Delta}^{-1} \mathbf{N}_{13} & = 
	\begin{bmatrix}
	K+1 & 1 \\
	1 & K+1
	\end{bmatrix}, \\[1em]
	\mathbf{N}_{14}^\top \bm{\Delta}^{-1} \mathbf{N}_{14} & = 
	\begin{bmatrix}
	K+1 & -1 \\
	-1 & K+1
	\end{bmatrix}, \\[1em]
	\mathbf{N}_{13}^\top \bm{\Delta}^{-1} \mathbf{N}_{24} & = 
	\begin{bmatrix}
	K-1 & 1 \\
	1 & K-1
	\end{bmatrix}, \\[1em]
	\mathbf{N}_{13}^\top \bm{\Delta}_T^{-1} \mathbf{N}_{13} & = 
	\frac{2}{n_{13}}
	\begin{bmatrix}
	\phi_{b}+K \phi_{b\beta} & \phi_{b} \\
	\phi_{b} & \phi_{b}+K \phi_{b\beta}
	\end{bmatrix}, \\[1em]
	\mathbf{N}_{24}^\top \bm{\Delta}_T^{-1} \mathbf{N}_{24} & = 
	\frac{2}{n_{24}}
	\begin{bmatrix}
	K\phi_{b}+\phi_{b\beta} & \phi_{b\beta} \\
	\phi_{b\beta} & K\phi_{b}+\phi_{b\beta}
	\end{bmatrix}, \\[1em]
	\mathbf{N}_{13}^\top \bm{\Delta}_T^{-1} \mathbf{N}_{24} & = 
	\frac{1}{c_0}
	\begin{bmatrix}
	K-1 & -1 \\
	-1 & K-1
	\end{bmatrix}
	,
	\end{split}
	\end{equation}
	where $ c_0, \phi_{b}, \phi_{b\beta} $ are defined as in Eq.~\eqref{eq:appB-1} and 
	\begin{equation}
	\begin{split}
	n_{13} & = 2\phi_{b}^2 + \beta^2(1-\beta)^2+3\beta\phi_{b}(1-2b-\beta) \\
	& \hspace{1.5em} -4b\beta^2(1-b-\beta), \\
	n_{24} & = \phi_{b}(\phi_{b}+\phi_{b\beta}).
	\end{split}
	\end{equation}
	
	By the Sherman-Morrison-Woodbury formula~\cite{Horn2012}, we have
	\begin{equation}
	\begin{split}
	\mathbb{E} \left[ g_{\frac{1}{2}}(\bm{\nu}_1, \bm{\nu}_2, \bm{\nu}) \bm{\nu} \bm{\nu}^\top \right]^{-1} 
	& = \frac{1}{c_0} \bm{\Delta}^{-1} - \frac{1}{c_0^2} \bm{\Delta}^{-1} \mathbf{M}_{12} \bm{\Delta}^{-1}, \\[1em]
	\mathbb{E} \left[ g_{\frac{1}{2}}(\bm{\nu}_1, \bm{\nu}_3, \bm{\nu}) \bm{\nu} \bm{\nu}^\top \right]^{-1} 
	& = \bm{\Delta}_T^{-1} - \bm{\Delta}_T^{-1} \mathbf{M}_{13} \bm{\Delta}_T^{-1} \\
	& \hspace{1em} - \bm{\Delta}_T^{-1} \mathbf{M}_{24} \bm{\Delta}_T^{-1} \\
	& \hspace{1em} + \bm{\Delta}_T^{-1} \mathbf{M}_{24} \bm{\Delta}_T^{-1} \mathbf{M}_{13} \bm{\Delta}_T^{-1} \\
	& \hspace{1em} + \bm{\Delta}_T^{-1} \mathbf{M}_{13} \bm{\Delta}_T^{-1} \mathbf{M}_{24} \bm{\Delta}_T^{-1} \\
	& \hspace{1em} - \bm{\Delta}_T^{-1} \mathbf{M}_{13} \bm{\Delta}_T^{-1} \mathbf{M}_{24} \\
	& \hspace{2.4em} \bm{\Delta}_T^{-1} \mathbf{M}_{13} \bm{\Delta}_T^{-1}, \\[1em]
	\mathbb{E} \left[ g_{\frac{1}{2}}(\bm{\nu}_1, \bm{\nu}_4, \bm{\nu}) \bm{\nu} \bm{\nu}^\top \right]^{-1} 
	& = \frac{1}{c_0} \bm{\Delta}^{-1} - \frac{1}{c_0^2} \bm{\Delta}^{-1} \mathbf{M}_{14} \bm{\Delta}^{-1} \\
	& \hspace{1em} - \frac{1}{c_0^2} \bm{\Delta}^{-1} \mathbf{M}_{23}\bm{\Delta}^{-1} \\
	& \hspace{1em} + \frac{1}{c_0^3} \bm{\Delta}^{-1} \mathbf{M}_{23} \bm{\Delta}^{-1} \mathbf{M}_{14} \bm{\Delta}^{-1} \\
	& \hspace{1em} + \frac{1}{c_0^3} \bm{\Delta}^{-1} \mathbf{M}_{14} \bm{\Delta}^{-1} \mathbf{M}_{23} \bm{\Delta}^{-1} \\
	& \hspace{1em} - \frac{1}{c_0^4} \bm{\Delta}^{-1} \mathbf{M}_{14} \bm{\Delta}^{-1} \mathbf{M}_{23} \\
	& \hspace{3.5em} \bm{\Delta}^{-1} \mathbf{M}_{14} \bm{\Delta}^{-1},
	\end{split}
	\end{equation}
	where $ c_0, c_{12}, c_{13}, c_{14}, c_{23}, c_{24} $ are defined as in Eq.~\eqref{eq:appB-1} and 
	\begin{equation}
	\begin{split}
	\mathbf{D}_{12} & = \frac{1}{c_{12}} \mathbf{I}_2 + \frac{1}{c_0} \mathbf{N}_{12}^\top \bm{\Delta}^{-1} \mathbf{N}_{12} 
	, \\
	\mathbf{D}_{13} & = \frac{1}{c_{13}} \mathbf{I}_2 + \mathbf{N}_{13}^\top \bm{\Delta}_T^{-1} \mathbf{N}_{13} 
	, \\
	\mathbf{D}_{14} & = \frac{1}{c_{14}} \mathbf{I}_2 + \frac{1}{c_0} \mathbf{N}_{14}^\top \bm{\Delta}^{-1} \mathbf{N}_{14} 
	, \\
	\mathbf{M}_{12} & = \mathbf{N}_{12} \mathbf{D}_{12}^{-1} \mathbf{N}_{12}^\top, \\
	\mathbf{M}_{13} & = \mathbf{N}_{13} \mathbf{D}_{13}^{-1} \mathbf{N}_{13}^\top, \\
	\mathbf{M}_{14} & = \mathbf{N}_{14} \mathbf{D}_{14}^{-1} \mathbf{N}_{14}^\top, \\
	\mathbf{D}_{23} & = \frac{1}{c_{23}} \mathbf{I}_2 + \frac{1}{c_0} \mathbf{N}_{23}^\top \bm{\Delta}^{-1} \mathbf{N}_{23} \\
	& \hspace{1.5em} - \frac{1}{c_0^2} \mathbf{N}_{23}^\top \bm{\Delta}^{-1} \mathbf{M}_{14} \bm{\Delta}^{-1} \mathbf{N}_{23}, \\
	\mathbf{D}_{24} & = \frac{1}{c_{24}} \mathbf{I}_2 + \mathbf{N}_{24}^\top \bm{\Delta}_T^{-1} \mathbf{N}_{24} \\
	& \hspace{1.5em} - \mathbf{N}_{24}^\top \bm{\Delta}_T^{-1} \mathbf{M}_{13} \bm{\Delta}_T^{-1} \mathbf{N}_{24}, \\ 
	\mathbf{M}_{23} & = \mathbf{N}_{23} \mathbf{D}_{23}^{-1} \mathbf{N}_{23}^\top, \\
	\mathbf{M}_{24} & = \mathbf{N}_{24} \mathbf{D}_{24}^{-1} \mathbf{N}_{24}^\top.
	\end{split}
	\end{equation}
	
	Again by canonical latent position matrix $ \bm{\nu}_Z(K,2K) $ as in Eq.~\eqref{eq:nuZK2K}, we have
	\begin{equation}
	\begin{split}
	\left(\bm{\nu}_1 - \bm{\nu}_2 \right)^\top \bm{\Delta} \left(\bm{\nu}_1 - \bm{\nu}_2 \right) & = \beta^2, \\
	\left(\bm{\nu}_1 - \bm{\nu}_3 \right)^\top \bm{\Delta} \left(\bm{\nu}_1 - \bm{\nu}_3 \right) & = \frac{2}{K} (a-b)^2, \\
	\left(\bm{\nu}_1 - \bm{\nu}_4 \right)^\top \bm{\Delta} \left(\bm{\nu}_1 - \bm{\nu}_4 \right) & = \frac{2}{K} (a-b)^2 + \beta^2, \\
	\left(\bm{\nu}_1 - \bm{\nu}_3 \right)^\top \bm{\Delta} \bm{\Delta}_T^{-1} \bm{\Delta} \left(\bm{\nu}_1 - \bm{\nu}_3 \right) & = \frac{1}{c_0} \frac{2}{K} (a-b)^2.
	\end{split}
	\end{equation}
	
	Similarly, we have
	\begin{equation}
	\begin{split}
	\mathbf{N}_{12}^\top  \left(\bm{\nu}_1 - \bm{\nu}_2 \right) & = \beta 
	\begin{bmatrix}
	1 & -1
	\end{bmatrix}^\top, \\
	\mathbf{N}_{13}^\top  \left(\bm{\nu}_1 - \bm{\nu}_3 \right) & = (a-b)
	\begin{bmatrix}
	1 & -1
	\end{bmatrix}^\top, \\
	\mathbf{N}_{14}^\top \left(\bm{\nu}_1 - \bm{\nu}_4 \right) & = (a-b+\beta)
	\begin{bmatrix}
	1 & -1
	\end{bmatrix}^\top, \\
	\mathbf{N}_{23}^\top \left(\bm{\nu}_1 - \bm{\nu}_4 \right) & = (a-b-\beta)
	\begin{bmatrix}
	1 & -1
	\end{bmatrix}^\top, \\
	\mathbf{N}_{13}^\top \bm{\Delta}_T^{-1} \bm{\Delta} \left(\bm{\nu}_1 - \bm{\nu}_3 \right) & = \frac{(a-b)}{c_0}
	\begin{bmatrix}
	1 & -1
	\end{bmatrix}^\top
	.
	\end{split}
	\end{equation}
	
	Then with all the results above, we have
	\begin{equation}
	\label{eq:C121314}
	\begin{split}
	C_{12} 
	& = \frac{K \beta^2}{2 D_4}, \\[1em]
	C_{13} 
	& = \frac{(a-b)^2}{K (\phi_a + \phi_b + \phi_{\beta})}, \\[1em]
	C_{14}
	& = \frac{K^2 \beta^2 (\phi_a+\phi_b+\phi_{\beta})+2K N_3+4N_4}{2K[2(\phi_a^2-\phi_b^2)+D_5]},
	\end{split}
	\end{equation}
	where $ \phi_{a}, \phi_{b} $ are defined as in Eq.~\eqref{eq:appB-1} and 
	\begin{equation}
	\label{eq:appB-2}
	\begin{split}
	\phi_{\beta} & = \beta (1 - a - b - \beta), \\
	D_3 & = K - 2a - 2(K-1)b - K\beta, \\
	D_4 & = 2 \phi_a + 2(K-1) \phi_b + \beta D_3, \\
	N_3 & = (a-b)^2 [2\phi_b+\beta(1+\beta-2b)], \\
	N_4 & = (a-b)^3(1-a-b-\beta), \\
	D_5 & = 2\beta(a-b)[(1-a-b-\beta)-2(\phi_a+\phi_b) \\
	& \hspace{1em} -\phi_{\beta}+2b(a+\beta)]+K \{2\phi_b(\phi_a+\phi_b) \\
	& \hspace{1em} - 2b\beta(\phi_b+a-b^2)-2ab\phi_{\beta}+\beta(1-\beta)[\phi_a \\
	& \hspace{1em} +(3b+\beta)(1-\beta)-a\beta-5b^2] \}.
	\end{split}
	\end{equation}
	
	Then we have the approximate Chernoff information for Algorithm~1 as
	\begin{equation}
	\rho_1^* \approx \min_{\ell \in \{2, 3, 4 \}} C_{1\ell},
	\end{equation}
	where $ C_{1\ell} $ for $ \ell \in \{2, 3, 4 \} $ are defined as in Eq.~\eqref{eq:C121314}. Also observe that
	\begin{equation}
	\begin{split}
	C_{12} - C_{14} & = \frac{-(a-b)^2 N_6^2}{K D_4 [2(\phi_a^2-\phi_b^2)+D_5]}, \\
	C_{13} - C_{14} & = \frac{- \beta^2[2(a-b)^2+K(\phi_a + \phi_b + \phi_{\beta})]^2}{2K (\phi_a + \phi_b + \phi_{\beta}) [2(\phi_a^2-\phi_b^2)+D_5]},
	\end{split}
	\end{equation}
	where $ \phi_{a}, \phi_{b} $ are defined as in Eq.~\eqref{eq:appB-1}, $ \phi_{\beta}, D_4, D_5 $ are defined as in Eq.~\eqref{eq:appB-2} and 
	\begin{equation}
	\begin{split}
	N_5 & = \beta[K-2a-2(K-1)b], \\
	N_6 & = 2\phi_a +2(K-1) \phi_b + N_5.
	\end{split}
	\end{equation}
	
	Subsequent calculations and simplification yield $ \rho_1^* $ as
	\begin{equation}
	\label{eq:rho1star}
	\rho_1^* \approx
	\begin{cases}
	\frac{K \beta^2}{2 D_4} & \text{if} \;\; \delta \leq 0  \\[1em]
	\frac{(a-b)^2}{K (\phi_a + \phi_b + \phi_{\beta})} & \text{if} \;\; \delta > 0
	\end{cases},
	\end{equation}
	where $ \phi_a, \phi_b, \phi_{\beta} $ are defined as in Eq.~\eqref{eq:phiabbeta} and 
	\begin{equation}
	\label{eq:condition}
	\begin{split}
	D_3 & = K - 2a - 2(K-1)b - K\beta, \\
	D_4 & = 2 \phi_a + 2(K-1) \phi_b + \beta D_3, \\
	\delta & = K^2 \beta^2 (\phi_a + \phi_b + \phi_{\beta}) - 2(a-b)^2 D_4.
	\end{split}
	\end{equation}
	
	Again by the same derivations~\cite{Cape2019}, we have the approximate Chernoff information for Algorithm~2 as
	\begin{equation}
	\rho_2^* \approx \frac{(a-b)^2}{K \left[a(1-a) + b(1-b) \right]} = \frac{(a-b)^2}{K (\phi_a + \phi_b)},
	\end{equation}
	where $ \phi_a $ and $ \phi_b $ are defined as in Eq.~\eqref{eq:phiabbeta}.
\end{proof}

\section{Simulation Details}

\label{app:B}

The implementation of our algorithms can be found at https://github.com/CongM/sbm-cov. 

Table~\ref{tab:fig4rp} summarizes the detailed results associated with Fig.~4 right panel for correspondence between Chernoff analysis and simulations. Table~\ref{tab:fig5} summarizes the detailed results associated with Fig.~5 for two-block rank one model with one five-categorical covariate. Table~\ref{tab:fig6} summarizes the detailed results associated with Fig.~6 for two-block homogeneous model with one five-categorical covariate.

\begin{table*}
	\renewcommand{\arraystretch}{1.3} 
	\caption{Detailed results associated with Fig.~4 right panel} 
	\label{tab:fig4rp}
	\centering
	\begin{threeparttable}
	\begin{tabular}{cccccccccc}
		\hline \hline
		$n$ \tnote{1} & $p$ \tnote{2}  & $q$ \tnote{2}    & $ \beta $ \tnote{3} & $ \widehat{\beta} \tnote{4} $ & ARI (Algo~1) \tnote{5}  & ARI (Algo~2 with $ \beta $) \tnote{5}  & ARI (Algo~2 with $ \widehat{\beta} $) \tnote{5} & Elapsed (Algo~1) \tnote{6}  & Elapsed (Algo~2) \tnote{6} \\ \hline \hline 
		100 & 0.3 & 0.668 & 0.49 &  0.489 & 0.858 ($ \pm $ 0.025) & 0.951 ($ \pm $ 0.005) & \textbf{0.952 ($ \pm $ 0.005)} & 0.032 & 0.155 \\
		140 & 0.3 & 0.668 & 0.49 &  0.487 & 0.957 ($ \pm $ 0.013) & \textbf{0.983 ($ \pm $ 0.002)} & \textbf{0.983 ($ \pm $ 0.002)} & 0.036 & 0.263 \\
		180 & 0.3 & 0.668 & 0.49 &  0.488 & 0.980 ($ \pm $ 0.010) & \textbf{0.995 ($ \pm $ 0.000)} & \textbf{0.995 ($ \pm $ 0.000)} & 0.040 & 0.402 \\
		220 & 0.3 & 0.668 & 0.49 & 0.489  & 0.997 ($ \pm $ 0.000) & \textbf{0.998 ($ \pm $ 0.000)} & \textbf{0.998 ($ \pm $ 0.000)} & 0.052 & 0.592 \\
		260 & 0.3 & 0.668 & 0.49 & 0.489  & \textbf{0.999 ($ \pm $ 0.000)} & \textbf{0.999 ($ \pm $ 0.000)} & \textbf{0.999 ($ \pm $ 0.000)} & 0.063 & 0.787 \\
		\hline
		100 & 0.3 & 0.564 & 0.49 &  0.478 & 0.291 ($ \pm $ 0.030) & 0.522 ($ \pm $ 0.027) & \textbf{0.545 ($ \pm $ 0.026)} & 0.039 & 0.160 \\
		140 & 0.3 & 0.564 & 0.49 &  0.485 & 0.572 ($ \pm $ 0.034) & \textbf{0.783 ($ \pm $ 0.010)} & 0.771 ($ \pm $ 0.013) & 0.049 & 0.280 \\
		180 & 0.3 & 0.564 & 0.49 &  0.486 & 0.783 ($ \pm $ 0.025) & 0.873 ($ \pm $ 0.004) & \textbf{0.874 ($ \pm $ 0.005)} & 0.056 & 0.446 \\
		220 & 0.3 & 0.564 & 0.49 & 0.490  & 0.874 ($ \pm $ 0.016) & \textbf{0.921 ($ \pm $ 0.003)} & 0.920 ($ \pm $ 0.003) & 0.065 & 0.631 \\
		260 & 0.3 & 0.564 & 0.49 & 0.489  & 0.905 ($ \pm $ 0.018) & \textbf{0.949 ($ \pm $ 0.003)} & \textbf{0.949 ($ \pm $ 0.003)} & 0.076 & 0.870 \\ \hline \hline
	\end{tabular}
	\begin{tablenotes}
		\item[1] Number of vertices in the simulated adjacency matrices.
		\item[2] Adjacency matrices are simulated from a two-block SBM where vertices within block 1 connect with probability $ p^2 $, vertices within block 2 connect with probability $ q^2 $, and vertices across two blocks connect with probability $ pq $.
		\item[3] Vertex covariate effect, see Definition 3 and Remark 1 for details. 
		\item[4] Estimated vertex covariate effect by Algorithm~2.
		\item[5] Reported as \texttt{mean($\pm$stderr)} from 100 trials. Best results in bold.
		\item[6] Reported as \texttt{mean} in seconds from 100 trials. 
	\end{tablenotes}
    \end{threeparttable}
\end{table*}

\begin{table*}
	\renewcommand{\arraystretch}{1.3} 
	\caption{Detailed results associated with Fig.~5} 
	\label{tab:fig5}
	\centering
	\begin{threeparttable}
	\begin{tabular}{cccccccccc}
		\hline \hline
		$n$ \tnote{1} & $p$ \tnote{2}  & $q$  \tnote{2}   & $ \beta $ \tnote{3} & $ \widehat{\beta} $ \tnote{4} & ARI (Algo~1) \tnote{5}  & ARI (Algo~2 with $ \beta $)  \tnote{5} & ARI (Algo~2 with $ \widehat{\beta} $) \tnote{5} & Elapsed (Algo~1) \tnote{6}  & Elapsed (Algo~2) \tnote{6}  \\ \hline \hline
		2000 & 0.3 & 0.350 & 0.4 & 0.401  & 0.013 ($ \pm $ 0.000) & 0.388 ($ \pm $ 0.012) & \textbf{0.395 ($ \pm $ 0.010)} & 7.133 & 83.225 \\
		2000 & 0.3 & 0.375 & 0.4 & 0.400  & 0.283 ($ \pm $ 0.030) & \textbf{0.763 ($ \pm $ 0.002)} & \textbf{0.763 ($ \pm $ 0.002)} & 6.476 & 77.849 \\
		2000 & 0.3 & 0.400 & 0.4 & 0.399  & 0.855 ($ \pm $ 0.014) & \textbf{0.931 ($ \pm $ 0.000)} & \textbf{0.931 ($ \pm $ 0.000)} & 6.268 & 76.409 \\
		2000 & 0.3 & 0.425 & 0.4 & 0.399  & 0.967 ($ \pm $ 0.007) & \textbf{0.985 ($ \pm $ 0.000)} & \textbf{0.985 ($ \pm $ 0.000)} & 5.854 & 74.726 \\
		2000 & 0.3 & 0.450 & 0.4 & 0.399  & 0.990 ($ \pm $ 0.005) & \textbf{0.998 ($ \pm $ 0.000)} & \textbf{0.998 ($ \pm $ 0.000)} & 5.793 & 76.510 \\ 
		\hline
		2000 & 0.3 & 0.375 & 0.10 & 0.097  & 0.774 ($ \pm $ 0.007) & \textbf{0.803 ($ \pm $ 0.002)} & 0.802 ($ \pm $ 0.002) & 5.890 & 75.155 \\
		2000 & 0.3 & 0.375 & 0.15 & 0.149  & 0.636 ($ \pm $ 0.026) & 0.789 ($ \pm $ 0.001) & \textbf{0.790 ($ \pm $ 0.001)} & 6.086 & 75.391 \\
		2000 & 0.3 & 0.375 & 0.20 & 0.199  & 0.452 ($ \pm $ 0.036) & 0.779 ($ \pm $ 0.002) & \textbf{0.780 ($ \pm $ 0.002)} & 6.345 & 75.126 \\
		2000 & 0.3 & 0.375 & 0.25 & 0.249  & 0.465 ($ \pm $ 0.034) & \textbf{0.770 ($ \pm $ 0.002)} & \textbf{0.770 ($ \pm $ 0.002)} & 6.198 & 74.796 \\
		2000 & 0.3 & 0.375 & 0.30 & 0.300  & 0.340 ($ \pm $ 0.034) & 0.766 ($ \pm $ 0.002) & \textbf{0.767 ($ \pm $ 0.001)} & 6.232 & 73.747 \\ \hline 	\hline
	\end{tabular}
	\begin{tablenotes}
		\item[1] Number of vertices in the simulated adjacency matrices.
		\item[2] Adjacency matrices are simulated from a two-block SBM where vertices within block 1 connect with probability $ p^2 $, vertices within block 2 connect with probability $ q^2 $, and vertices across two blocks connect with probability $ pq $.
		\item[3] Vertex covariate effect, see Definition 3 and Remark 1 for details. 
		\item[4] Estimated vertex covariate effect by Algorithm~2.
		\item[5] Reported as \texttt{mean($\pm$stderr)} from 100 trials. Best results in bold.
		\item[6] Reported as \texttt{mean} in seconds from 100 trials. 
	\end{tablenotes}
\end{threeparttable}
\end{table*}

\begin{table*}
	\renewcommand{\arraystretch}{1.3} 
	\caption{Detailed results associated with Fig.~6} 
	\label{tab:fig6}
	\centering
	\begin{threeparttable}
	\begin{tabular}{cccccccccc}
		\hline \hline
		$n$ \tnote{1} & $a$ \tnote{2}  & $b$  \tnote{2}   & $ \beta $ \tnote{3} & $ \widehat{\beta} $ \tnote{4} & ARI (Algo~1) \tnote{5}  & ARI (Algo~2 with $ \beta $)  \tnote{5} & ARI (Algo~2 with $ \widehat{\beta} $) \tnote{5} & Elapsed (Algo~1) \tnote{6}  & Elapsed (Algo~2) \tnote{6}  \\ \hline \hline
		2000 & 0.120 & 0.1 & 0.2 & 0.201 & 0.000 ($ \pm $ 0.000) & 0.228 ($ \pm $ 0.014) & \textbf{0.230 ($ \pm $ 0.014)} & 6.608 & 73.482 \\
		2000 & 0.125 & 0.1 & 0.2 & 0.200  & 0.001 ($ \pm $ 0.000) & \textbf{0.608 ($ \pm $ 0.006)} & 0.583 ($ \pm $ 0.013) & 6.538 & 74.862 \\
		2000 & 0.130 & 0.1 & 0.2 & 0.200  & 0.039 ($ \pm $ 0.007) & \textbf{0.790 ($ \pm $ 0.008)} & 0.789 ($ \pm $ 0.008) & 6.138 & 75.069 \\
		2000 & 0.135 & 0.1 & 0.2 & 0.199  & 0.058 ($ \pm $ 0.008) & \textbf{0.893 ($ \pm $ 0.009)} & \textbf{0.893 ($ \pm $ 0.009)} & 5.649 & 71.505 \\
		2000 & 0.140 & 0.1 & 0.2 & 0.199  & 0.075 ($ \pm $ 0.010) & \textbf{0.954 ($ \pm $ 0.000)} & \textbf{0.954 ($ \pm $ 0.000)} & 5.497 & 72.871 \\
		\hline
		2000 & 0.135 & 0.1 & -0.09 &  $-0.088$ & 0.098 ($ \pm $ 0.015) & \textbf{0.969 ($ \pm $ 0.000)} & \textbf{0.969 ($ \pm $ 0.000)} & 5.448 & 71.333 \\
		2000 & 0.135 & 0.1 & -0.08 &  $-0.077$ & 0.088 ($ \pm $ 0.012) & \textbf{0.967 ($ \pm $ 0.000)} & \textbf{0.967 ($ \pm $ 0.000)} & 5.624 & 71.421 \\
		2000 & 0.135 & 0.1 & -0.07 &  $-0.065$ & 0.073 ($ \pm $ 0.011) & \textbf{0.965 ($ \pm $ 0.000)} & \textbf{0.965 ($ \pm $ 0.000)} & 5.498 & 72.388 \\
		2000 & 0.135 & 0.1 & -0.06 & $-0.052$ & 0.083 ($ \pm $ 0.013) & \textbf{0.963 ($ \pm $ 0.000)} & \textbf{0.963 ($ \pm $ 0.000)} & 5.583 & 72.096 \\
		2000 & 0.135 & 0.1 & -0.05 & $-0.037$  & 0.069 ($ \pm $ 0.008) & \textbf{0.960 ($ \pm $ 0.000)} & \textbf{0.960 ($ \pm $ 0.000)} & 6.819 & 74.507 \\ 
		\hline \hline 
	\end{tabular}
	\begin{tablenotes}
		\item[1] Number of vertices in the simulated adjacency matrices.
		\item[2] Adjacency matrices are simulated from a two-block SBM where vertices from the same block connect with probability $ a $, and vertices across two blocks connect with probability $ b $.
		\item[3] Vertex covariate effect, see Definition 3 and Remark 1 for details. 
		\item[4] Estimated vertex covariate effect by Algorithm~2.
		\item[5] Reported as \texttt{mean($\pm$stderr)} from 100 trials. Best results in bold.
		\item[6] Reported as \texttt{mean} in seconds from 100 trials. 
	\end{tablenotes}
\end{threeparttable}
\end{table*}



\ifCLASSOPTIONcaptionsoff
  \newpage
\fi



\bibliographystyle{IEEEtran}
\bibliography{MMHCAP20_IEEE.bib}
%



%

\begin{IEEEbiography}[{\includegraphics[width=1in,height=1.25in,clip,keepaspectratio]{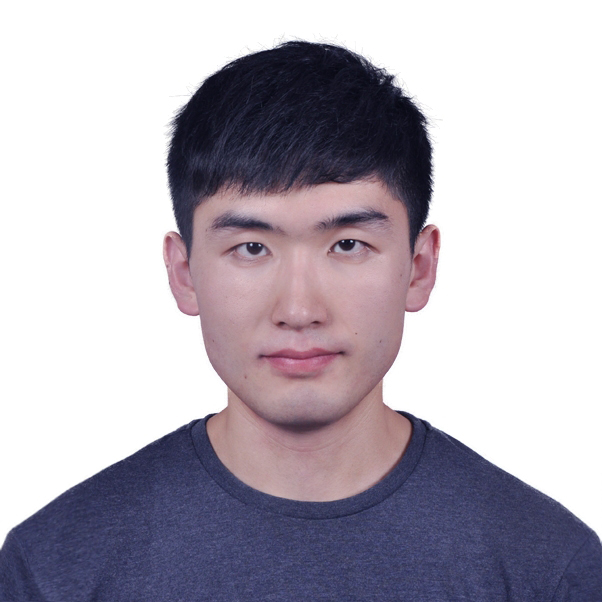}}]{Cong Mu}
received the BS degree in statistics from Sun Yat-Sen University, in 2017, the MSE degree in applied mathematics and statistics from Johns Hopkins University, in 2019. He is currently working toward the PhD degree in applied mathematics and statistics from Johns Hopkins University. His research interests include high-dimensional and graph inference, and computer vision.
\end{IEEEbiography}

\begin{IEEEbiography}[{\includegraphics[width=1in,height=1.25in,clip,keepaspectratio]{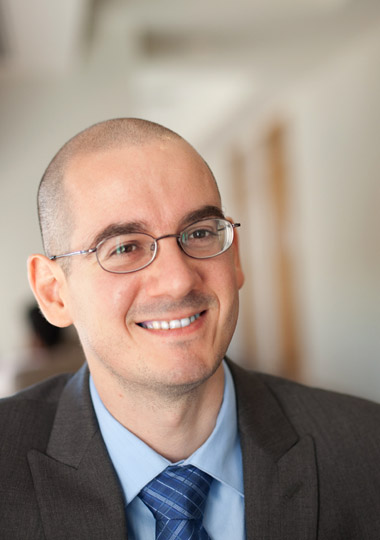}}]{Angelo Mele}
received the BA (Laurea) degree in Economics from Bocconi University, Italy in 2002, the MA in Economics from New York University in 2005 and the PhD degree in Economics from University of Illinois at Urbana-Champaign in 2011. Currently he is an Associate Professor of Economics at Johns Hopkins University Carey Business School. His research interests include the econometrics of network models, economics of social interactions, racial segregation and homophily, online contagion and computational methods.
\end{IEEEbiography}

\begin{IEEEbiography}[{\includegraphics[width=1in,height=1.25in,clip,keepaspectratio]{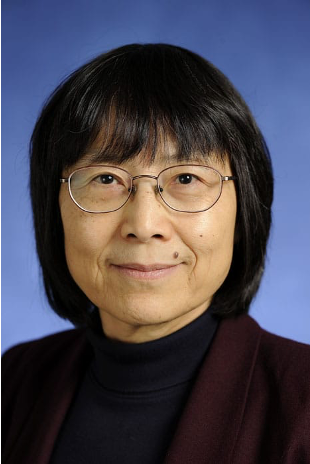}}]{Lingxin Hao}
	received the BA degree in English, South China Normal University, China in 1982, the MA degree in Sociology, Sun Yat-sen University, China, in 1985, and Ph.D. in Sociology, the University of Chicago, in 1990. Before joining the Johns Hopkins University in 1996, she was a Postdoc Fellow at the Labor and Population Program, RAND, and Assistant-to-Associate Professor at the University of Iowa. Currently she is Professor of Sociology and Director of Hopkins Population Center. Dr. Hao has been the principle investigator for several multi-year projects supported by federal grants from NIH and NSF. She was a Residential Fellow at the Russell Sage Foundation in 2002-2003 and a Residential Fellow at the Spenser Foundation in 2007. Her research interests include social demography, social inequality, migration, family and public policy, sociology of education, and quantitative and computational methods.
\end{IEEEbiography}

\begin{IEEEbiography}[{\includegraphics[width=1in,height=1.25in,clip,keepaspectratio]{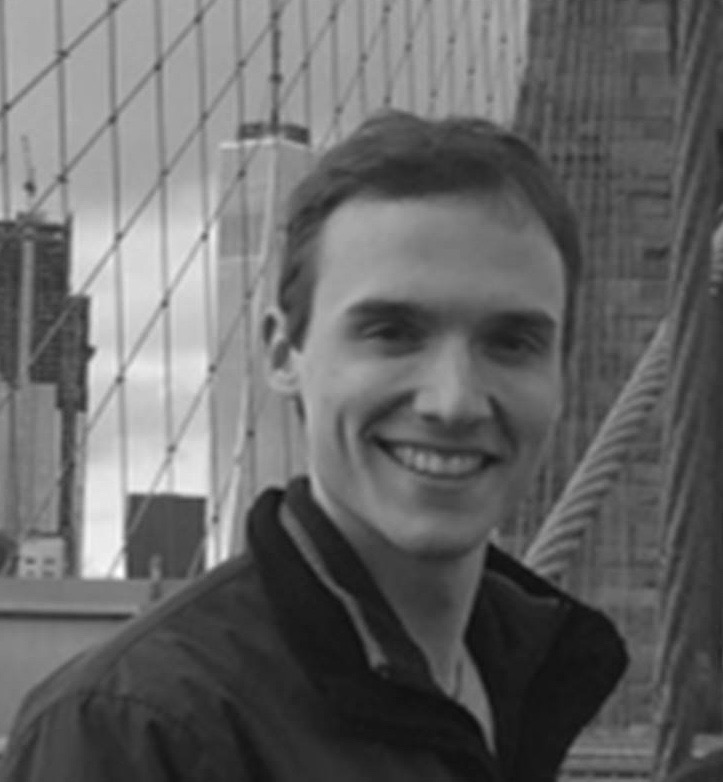}}]{Joshua Cape}
	received the BA degree in mathematics and economics from Rhodes College in 2014, the MSE degree in applied mathematics and statistics from Johns Hopkins University in 2016, and the PhD degree in applied mathematics and statistics from Johns Hopkins University in 2019. Previously, he was a National Science Foundation Postdoctoral Research Fellow in the Department of Statistics at the University of Michigan between 2019 and 2020. Currently, he is an Assistant Professor of Statistics at the University of Pittsburgh. His research interests include statistical machine learning, multivariate analysis, networks, and matrix analysis. 
\end{IEEEbiography}

\begin{IEEEbiography}[{\includegraphics[width=1in,height=1.25in,clip,keepaspectratio]{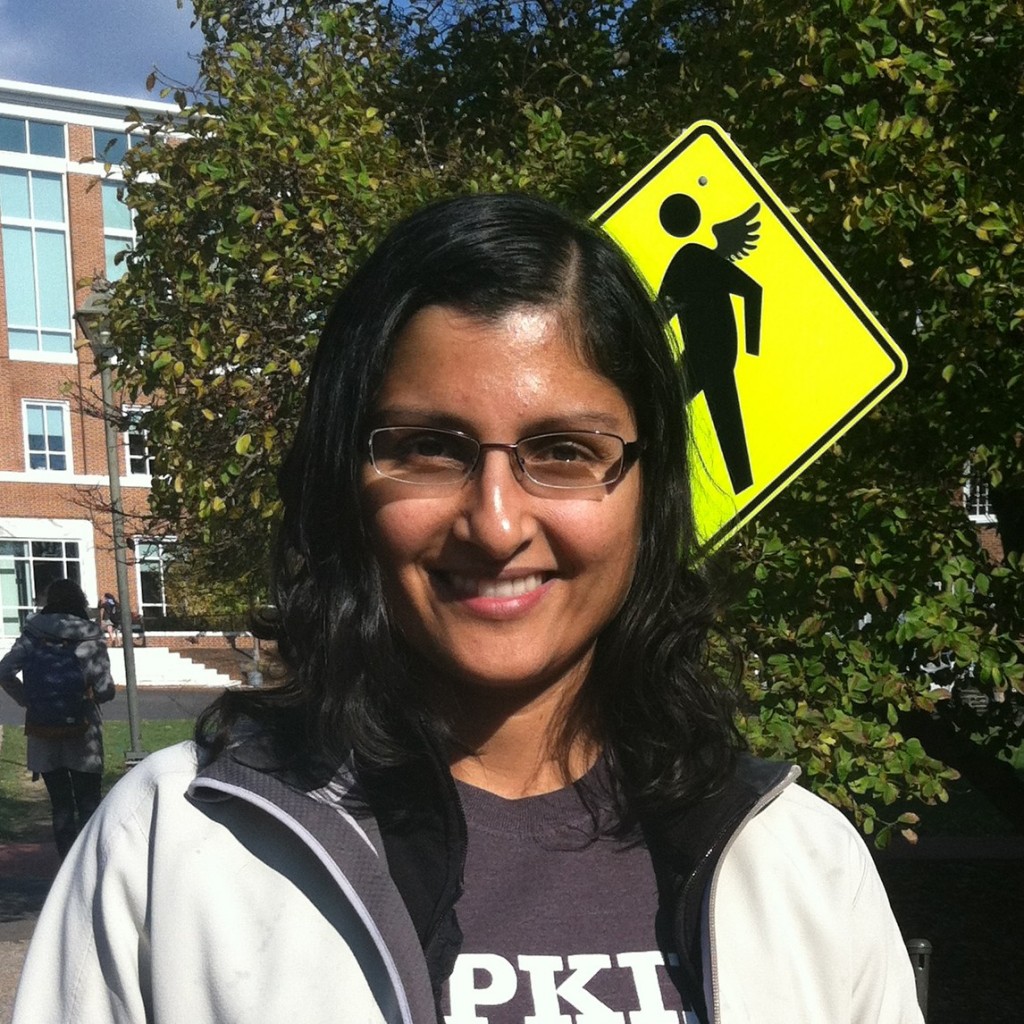}}]{Avanti Athreya}
	received the BS degree from Iowa State University, in 1997, the MS degree from the University of Washington, in 2000, and
	the PhD degree from the University of Maryland, College Park, in 2009. She was a visiting assistant professor with Duke University and a postdoctoral fellow with SAMSI prior to coming to Johns Hopkins University, in 2011, where she is currently an assistant research professor. Her research interests include random  graph inference, probability, and stochastic processes.
\end{IEEEbiography}

\begin{IEEEbiography}[{\includegraphics[width=1in,height=1.25in,clip,keepaspectratio]{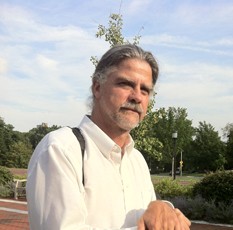}}]{Carey E. Priebe}
	received the BS degree in mathematics from Purdue University, in 1984, the MS degree in computer science from San Diego State University, in 1988, and the PhD degree in information technology (computational statistics) from George Mason University, in 1993. From 1985 to 1994, he worked as a mathematician and scientist in the US Navy research and development laboratory system. Since 1994, he has been a professor in the Department of Applied Mathematics and Statistics, Johns Hopkins University (JHU). At JHU, he holds joint appointments in the Department of Computer Science, Department of Electrical and Computer Engineering, Center for Imaging Science, Human Language Technology Center of Excellence, and Whitaker Biomedical Engineering Institute. His research interests include computational statistics, kernel and mixture estimates, statistical pattern recognition, statistical image analysis, dimensionality reduction, model selection, and statistical inference for high-dimensional and graph data. He is a lifetime member of the IMS, an elected member of the ISI, and a fellow of the ASA.
\end{IEEEbiography}







\end{document}